\documentclass[longauth]{aa}
\usepackage{graphicx}
\usepackage{txfonts,textcomp}
\usepackage{xcolor}
\usepackage{hyperref}
\hypersetup{
  colorlinks,
  linkcolor={red},
  citecolor={blue!80!black},
  urlcolor={blue!80!black}
}
\usepackage{color}
\usepackage{siunitx}
\usepackage{mathabx}
\newcommand{\ticA}{{\rm TOI-5678}}
\usepackage[normalem]{ulem}
\usepackage{orcidlink}

\begin{document}

\title{TOI-5678\,b: A 48-day transiting Neptune-mass planet characterized with CHEOPS and HARPS\thanks{Based on data from CHEOPS Prog ID: CH\_PR110048 and HARPS Prog ID: 108.22L8.001. Reduced CHEOPS and HARPS data are available at the Strasbourg astronomical Data Center (CDS).}}

\titlerunning{48-day transiting Neptune-mass planet}

\author{
  S.~Ulmer-Moll\orcidlink{0000-0003-2417-7006} \inst{\ref{geneva}, \ref{ inst2 }} \and
  H.~P.~Osborn\orcidlink{0000-0002-4047-4724} \inst{\ref{ inst2 }, \ref{ inst3 }} \and
  A.~Tuson\orcidlink{0000-0002-2830-9064} \inst{\ref{ inst4 }} \and
  J.~A.~Egger\orcidlink{0000-0003-1628-4231} \inst{\ref{ inst2 }} \and
  M.~Lendl\orcidlink{0000-0001-9699-1459} \inst{\ref{geneva}} \and
  P.~Maxted\orcidlink{0000-0003-3794-1317} \inst{\ref{ inst5 }} \and
  A.~Bekkelien \inst{\ref{geneva}} \and
  A.~E.~Simon\orcidlink{0000-0001-9773-2600} \inst{\ref{ inst2 }} \and
  G.~Olofsson\orcidlink{0000-0003-3747-7120} \inst{\ref{ inst6 }} \and
  V.~Adibekyan\orcidlink{0000-0002-0601-6199} \inst{\ref{ia-porto}} \and
  Y.~Alibert\orcidlink{0000-0002-4644-8818} \inst{\ref{ inst2 }} \and
  A.~Bonfanti\orcidlink{0000-0002-1916-5935} \inst{\ref{ inst8 }} \and
  F.~Bouchy \inst{\ref{geneva}} \and
  A.~Brandeker\orcidlink{0000-0002-7201-7536} \inst{\ref{ inst6 }} \and
  M.~Fridlund\orcidlink{0000-0002-0855-8426} \inst{\ref{ inst9 }, \ref{ inst10 }} \and
  D.~Gandolfi\orcidlink{0000-0001-8627-9628} \inst{\ref{ inst11 }} \and
  C.~Mordasini \inst{\ref{ inst2 }} \and
  C.~M.~Persson\orcidlink{0000-0003-1257-5146} \inst{\ref{ inst10 }} \and
  S.~Salmon\orcidlink{0000-0002-1714-3513} \inst{\ref{geneva}} \and
  L.~M.~Serrano\orcidlink{0000-0001-9211-3691} \inst{\ref{ inst11 }} \and
  S.~G.~Sousa\orcidlink{0000-0001-9047-2965} \inst{\ref{ia-porto}} \and
  T.~G.~Wilson\orcidlink{0000-0001-8749-1962} \inst{\ref{ inst13 }} \and
  M.~Rieder \inst{\ref{ inst2 }} \and
  J.~Hasiba \inst{\ref{ inst8 }} \and
  J.~Asquier \inst{\ref{ inst14 }} \and
  D.~Sicilia\orcidlink{0000-0001-7851-5168} \inst{\ref{ inst15 }} \and
  I.~Walter\orcidlink{0000-0002-5839-1521} \inst{\ref{ inst16 }} \and
  R.~Alonso\orcidlink{0000-0001-8462-8126} \inst{\ref{iac}, \ref{ inst18 }} \and
  G.~Anglada\orcidlink{0000-0002-3645-5977} \inst{\ref{ inst19 }, \ref{ inst20 }} \and
  D.~Barrado~y~Navascues\orcidlink{0000-0002-5971-9242} \inst{\ref{ inst21 }} \and
  S.~C.~C.~Barros\orcidlink{0000-0003-2434-3625} \inst{\ref{ia-porto}, \ref{ inst22 }} \and
  W.~Baumjohann\orcidlink{0000-0001-6271-0110} \inst{\ref{ inst8 }} \and
  M.~Beck\orcidlink{0000-0003-3926-0275} \inst{\ref{geneva}} \and
  T.~Beck \inst{\ref{ inst2 }} \and
  W.~Benz\orcidlink{0000-0001-7896-6479} \inst{\ref{ inst2 }, \ref{ inst23 }} \and
  N.~Billot\orcidlink{0000-0003-3429-3836} \inst{\ref{geneva}} \and
  X.~Bonfils\orcidlink{0000-0001-9003-8894} \inst{\ref{ inst24 }} \and
  L.~Borsato\orcidlink{0000-0003-0066-9268} \inst{\ref{ inst25 }} \and
  C.~Broeg\orcidlink{0000-0001-5132-2614} \inst{\ref{ inst2 }, \ref{ inst23 }} \and
  T.~Bárczy\orcidlink{0000-0002-7822-4413} \inst{\ref{ inst26 }} \and
  J.~Cabrera \inst{\ref{ inst27 }} \and
  S.~Charnoz\orcidlink{0000-0002-7442-491X} \inst{\ref{ inst28 }} \and
  M.~Cointepas \inst{\ref{geneva}, \ref{ inst24 }} \and
  A.~Collier~Cameron\orcidlink{0000-0002-8863-7828} \inst{\ref{ inst13 }} \and
  Sz.~Csizmadia\orcidlink{0000-0001-6803-9698} \inst{\ref{ inst27 }} \and
  P.~E.~Cubillos \inst{\ref{torino}, \ref{ inst8 }} \and
  M.~B.~Davies\orcidlink{0000-0001-6080-1190} \inst{\ref{ inst29 }} \and
  M.~Deleuil\orcidlink{0000-0001-6036-0225} \inst{\ref{ inst30 }} \and
  A.~Deline \inst{\ref{geneva}} \and
  L.~Delrez\orcidlink{0000-0001-6108-4808} \inst{\ref{ inst31 }, \ref{ inst32 }} \and
  O.~D.~S.~Demangeon\orcidlink{0000-0001-7918-0355} \inst{\ref{ia-porto}, \ref{ inst22 }} \and
  B.-O.~Demory\orcidlink{0000-0002-9355-5165} \inst{\ref{ inst23 }} \and
  X.~Dumusque \inst{\ref{geneva}} \and
  D.~Ehrenreich\orcidlink{0000-0001-9704-5405} \inst{\ref{geneva}, \ref{ inst33 }} \and
  N.L.~Eisner \inst{\ref{ny}} \and
  A.~Erikson \inst{\ref{ inst27 }} \and
  A.~Fortier\orcidlink{0000-0001-8450-3374} \inst{\ref{ inst2 }, \ref{ inst23 }} \and
  L.~Fossati\orcidlink{0000-0003-4426-9530} \inst{\ref{ inst8 }} \and
  M.~Gillon\orcidlink{0000-0003-1462-7739} \inst{\ref{ inst31 }} \and
  N.~Grieves\orcidlink{0000-0001-8105-0373} \inst{\ref{geneva}} \and
  M.~Güdel \inst{\ref{ inst34 }} \and
  J.~Hagelberg\orcidlink{0000-0002-1096-1433} \inst{\ref{geneva}} \and
  R.~Helled \inst{\ref{zurich}} \and
  S.~Hoyer\orcidlink{0000-0003-3477-2466} \inst{\ref{ inst30 }} \and
  K.~G.~Isaak\orcidlink{0000-0001-8585-1717} \inst{\ref{ inst35 }} \and
  L.~L.~Kiss \inst{\ref{ inst36 }, \ref{ inst37 }} \and
  J.~Laskar\orcidlink{0000-0003-2634-789X} \inst{\ref{ inst38 }} \and
  A.~{Lecavelier~des~Etangs}\orcidlink{0000-0002-5637-5253} \inst{\ref{ inst39 }} \and
  C.~Lovis\orcidlink{0000-0001-7120-5837} \inst{\ref{geneva}} \and
  D.~Magrin\orcidlink{0000-0003-0312-313X} \inst{\ref{ inst25 }} \and
  V.~Nascimbeni\orcidlink{0000-0001-9770-1214} \inst{\ref{ inst25 }} \and
  J.~Otegi \inst{\ref{geneva}} \and
  R.~Ottensammer \inst{\ref{ inst40 }} \and
  I.~Pagano\orcidlink{0000-0001-9573-4928} \inst{\ref{ inst15 }} \and
  E.~Pallé\orcidlink{0000-0003-0987-1593} \inst{\ref{iac}} \and
  G.~Peter\orcidlink{0000-0001-6101-2513} \inst{\ref{ inst41 }} \and
  G.~Piotto\orcidlink{0000-0002-9937-6387} \inst{\ref{ inst25 }, \ref{ inst42 }} \and
  D.~Pollacco \inst{\ref{ inst43 }} \and
  A.~Psaridi \inst{\ref{geneva}} \and
  D.~Queloz\orcidlink{0000-0002-3012-0316} \inst{\ref{ inst44 }, \ref{ inst45 }} \and
  R.~Ragazzoni\orcidlink{0000-0002-7697-5555} \inst{\ref{ inst25 }, \ref{ inst42 }} \and
  N.~Rando \inst{\ref{ inst14 }} \and
  H.~Rauer\orcidlink{0000-0002-6510-1828} \inst{\ref{ inst27 }, \ref{ inst46 }} \and
  I.~Ribas\orcidlink{0000-0002-6689-0312} \inst{\ref{ inst19 }, \ref{ inst20 }} \and
  N.~C.~Santos\orcidlink{0000-0003-4422-2919} \inst{\ref{ia-porto}, \ref{ inst22 }} \and
  G.~Scandariato\orcidlink{0000-0003-2029-0626} \inst{\ref{ inst15 }} \and
  A.~M.~S.~Smith\orcidlink{0000-0002-2386-4341} \inst{\ref{ inst27 }} \and
  M.~Steller\orcidlink{0000-0003-2459-6155} \inst{\ref{ inst8 }} \and
  G.~M.~Szabó \inst{\ref{ inst47 }, \ref{ inst48 }} \and
  D.~Ségransan\orcidlink{0000-0003-2355-8034} \inst{\ref{geneva}} \and
  N.~Thomas \inst{\ref{ inst2 }} \and
  S.~Udry\orcidlink{0000-0001-7576-6236} \inst{\ref{geneva}} \and
  V.~Van~Grootel\orcidlink{0000-0003-2144-4316} \inst{\ref{ inst32 }} \and
  J.~Venturini \inst{\ref{geneva}} \and
  N.~A.~Walton\orcidlink{0000-0003-3983-8778} \inst{\ref{ inst49 }}
}

\authorrunning{Ulmer-Moll et al.}
\institute{
  Observatoire Astronomique de l'Université de Genève, Chemin Pegasi 51, CH-1290 Versoix, Switzerland \label{geneva} \and
  Physikalisches Institut, University of Bern, Gesellschaftsstrasse 6, 3012 Bern, Switzerland \label{ inst2 } \and
  Department of Physics and Kavli Institute for Astrophysics and Space Research, Massachusetts Institute of Technology, Cambridge, MA 02139, USA \label{ inst3 } \and
  Astrophysics Group, Cavendish Laboratory, University of Cambridge, J.J. Thomson Avenue, Cambridge CB3 0HE, UK \label{ inst4 } \and
  Astrophysics Group, Keele University, Staffordshire, ST5 5BG, United Kingdom \label{ inst5 } \and
  Department of Astronomy, Stockholm University, AlbaNova University Center, 10691 Stockholm, Sweden \label{ inst6 } \and
  Instituto de Astrofísica e Ciências do Espaço, Universidade do Porto, CAUP, Rua das Estrelas, 4150-762 Porto, Portugal \label{ia-porto} \and
  Space Research Institute, Austrian Academy of Sciences, Schmiedlstrasse 6, A-8042 Graz, Austria \label{ inst8 } \and
  Leiden Observatory, University of Leiden, PO Box 9513, 2300 RA Leiden, The Netherlands \label{ inst9 } \and
  Department of Space, Earth and Environment, Chalmers University of Technology, Onsala Space Observatory, 439 92 Onsala, Sweden \label{ inst10 } \and
  Dipartimento di Fisica, Universita degli Studi di Torino, via Pietro Giuria 1, I-10125, Torino, Italy \label{ inst11 } \and 
  Centre for Exoplanet Science, SUPA School of Physics and Astronomy, University of St Andrews, North Haugh, St Andrews KY16 9SS, UK \label{ inst13 } \and
  ESTEC, European Space Agency, 2201AZ, Noordwijk, NL \label{ inst14 } \and
  INAF - Osservatorio Astrofisico di Catania, Via S. Sofia 78, 95123 Catania, Italy \label{ inst15 } \and
  German Aerospace Center (DLR), Institute of Optical Sensor Systems Rutherfordstraße 2 12489 Berlin \label{ inst16 } \and
  Instituto de Astrofisica de Canarias, 38200 La Laguna, Tenerife, Spain \label{iac} \and
  Departamento de Astrofisica, Universidad de La Laguna, 38206 La Laguna, Tenerife, Spain \label{ inst18 } \and
  Institut de Ciencies de l'Espai (ICE, CSIC), Campus UAB, Can Magrans s/n, 08193 Bellaterra, Spain \label{ inst19 } \and
  Institut d'Estudis Espacials de Catalunya (IEEC), 08034 Barcelona, Spain \label{ inst20 } \and
  Depto. de Astrofisica, Centro de Astrobiologia (CSIC-INTA), ESAC campus, 28692 Villanueva de la Cañada (Madrid), Spain \label{ inst21 } \and
  Departamento de Física e Astronomia, Faculdade de Ciências, Universidade do Porto, Rua do Campo Alegre, 4169-007 Porto, Portugal \label{ inst22 } \and
  Center for Space and Habitability, University of Bern, Gesellschaftsstrasse 6, 3012, Bern, Switzerland \label{ inst23 } \and
  Université Grenoble Alpes, CNRS, IPAG, 38000 Grenoble, France \label{ inst24 } \and
  INAF, Osservatorio Astronomico di Padova, Vicolo dell'Osservatorio 5, 35122 Padova, Italy \label{ inst25 } \and
  Admatis, 5. Kandó Kálmán Street, 3534 Miskolc, Hungary \label{ inst26 } \and
  Institute of Planetary Research, German Aerospace Center (DLR), Rutherfordstrasse 2, 12489 Berlin, Germany \label{ inst27 } \and
  Université de Paris, Institut de physique du globe de Paris, CNRS, F-75005 Paris, France \label{ inst28 } \and
  INAF - Osservatorio Astrofisico di Torino, Strada Osservatorio, 20 I-10025 Pino Torinese (TO), Italy \label{torino} \and
  Centre for Mathematical Sciences, Lund University, Box 118, 221 00 Lund, Sweden \label{ inst29 } \and
  Aix Marseille Univ, CNRS, CNES, LAM, 38 rue Frédéric Joliot-Curie, 13388 Marseille, France \label{ inst30 } \and
  Astrobiology Research Unit, Université de Liège, Allée du 6 Août 19C, B-4000 Liège, Belgium \label{ inst31 } \and
  Space sciences, Technologies and Astrophysics Research (STAR) Institute, Université de Liège, Allée du 6 Août 19C, 4000 Liège, Belgium \label{ inst32 } \and
  Centre Vie dans l'Univers, Faculté des sciences, Université de Genève, Quai Ernest-Ansermet 30, CH-1211 Genève 4, Switzerland \label{ inst33 } \and
  Center for Computational Astrophysics, Flatiron Institute, New York, NY 10010, USA \label{ny} \and
  University of Vienna, Department of Astrophysics, Türkenschanzstrasse 17, 1180 Vienna, Austria \label{ inst34 } \and
  Institute for Computational Science, University of Zurich, Winterthurerstr. 190, CH- 8057 Zurich, Switzerland \label{zurich} \and
  Science and Operations Department - Science Division (SCI-SC), Directorate of Science, European Space Agency (ESA), European Space Research and Technology Centre (ESTEC), Keplerlaan 1, 2201-AZ Noordwijk, The Netherlands \label{ inst35 } \and
  Konkoly Observatory, Research Centre for Astronomy and Earth Sciences, 1121 Budapest, Konkoly Thege Miklós út 15-17, Hungary \label{ inst36 } \and
  ELTE Eötvös Loránd University, Institute of Physics, Pázmány Péter sétány 1/A, 1117 Budapest, Hungary \label{ inst37 } \and
  IMCCE, UMR8028 CNRS, Observatoire de Paris, PSL Univ., Sorbonne Univ., 77 av. Denfert-Rochereau, 75014 Paris, France \label{ inst38 } \and
  Institut d'astrophysique de Paris, UMR7095 CNRS, Université Pierre \& Marie Curie, 98bis blvd. Arago, 75014 Paris, France \label{ inst39 } \and
  Department of Astrophysics, University of Vienna, Tuerkenschanzstrasse 17, 1180 Vienna, Austria \label{ inst40 } \and
  Institute of Optical Sensor Systems, German Aerospace Center (DLR), Rutherfordstrasse 2, 12489 Berlin, Germany \label{ inst41 } \and
  Dipartimento di Fisica e Astronomia "Galileo Galilei", Universita degli Studi di Padova, Vicolo dell'Osservatorio 3, 35122 Padova, Italy \label{ inst42 } \and
  Department of Physics, University of Warwick, Gibbet Hill Road, Coventry CV4 7AL, UK \label{ inst43 } \and
  ETH Zurich, Department of Physics, Wolfgang-Pauli-Strasse 2, CH-8093 Zurich, Switzerland \label{ inst44 } \and
  Cavendish Laboratory, JJ Thomson Avenue, Cambridge CB3 0HE, UK \label{ inst45 } \and
  Zentrum für Astronomie und Astrophysik, Technische Universität Berlin, Hardenbergstr. 36, D-10623 Berlin, Germany \label{ inst46 } \and
  ELTE Eötvös Loránd University, Gothard Astrophysical Observatory, 9700 Szombathely, Szent Imre h. u. 112, Hungary \label{ inst47 } \and
  MTA-ELTE Exoplanet Research Group, 9700 Szombathely, Szent Imre h. u. 112, Hungary \label{ inst48 } \and
  Institute of Astronomy, University of Cambridge, Madingley Road, Cambridge, CB3 0HA, United Kingdom \label{ inst49 }
}

\date{Received 16 November 2022 / Accepted 17 April 2023}

\abstract
    {A large sample of long-period giant planets has been discovered thanks to long-term radial velocity surveys,
      but only a few dozen of these planets have a precise radius measurement.
      Transiting gas giants are crucial targets for the study of atmospheric composition across
      a wide range of equilibrium temperatures and, more importantly, for shedding light on the
      formation and evolution of planetary systems. Indeed, compared to hot Jupiters,
      the atmospheric properties and orbital parameters of cooler gas giants are unaltered by
      intense stellar irradiation and tidal effects.}
    {We aim to identify long-period planets in
      the Transiting Exoplanet Survey Satellite (TESS) data as single or duo-transit events. Our goal is to solve
      the orbital periods of TESS duo-transit candidates with the use of additional space-based
      photometric observations and to collect follow-up spectroscopic observations in
      order to confirm the planetary nature and measure the mass of the candidates.}
    {We use the CHaracterising ExOPlanet Satellite (CHEOPS) to observe the highest-probability
      period aliases in order to discard or confirm a transit event at a given period.
      Once a period is confirmed, we jointly model the TESS and CHEOPS light curves along with the radial velocity datasets
      to measure the orbital parameters of the system and obtain precise mass and radius measurements.
    }
    { We report the discovery of a long-period transiting Neptune-mass planet orbiting the G7-type star \ticA.
      Our spectroscopic analysis shows that \ticA\ is a star with a solar metallicity.
      The TESS light curve of \ticA\ presents two transit events separated by almost two years.
      In addition, CHEOPS observed the target as part of its Guaranteed Time Observation program.
      After four non-detections corresponding to possible periods, CHEOPS detected a transit event
      matching a unique period alias.
      Follow-up radial velocity observations were carried out with the ground-based high-resolution spectrographs CORALIE and HARPS.
      Joint modeling reveals that \ticA\ hosts a 47.73 day period planet, and we measure an orbital eccentricity consistent with zero at 2 $\sigma$.
      The planet \ticA\,b has a mass of $20\pm4$ Earth masses ($\rm M_{\oplus}$) and a radius of $\rm 4.91\pm0.08\,R_{\oplus}$.
      Using interior structure modeling, we find that \ticA\,b is composed of a low-mass core surrounded by
      a large H/He layer with a mass of $3.2^{+1.7}_{-1.3}$\,M$_{\oplus}$.}
    {\ticA\,b is part of a growing sample of well-characterized transiting gas giants receiving moderate amounts of stellar insolation ($11\,S_{\oplus}$).
      Precise density measurement gives us insight into their interior composition, and the objects orbiting bright stars
      are suitable targets to study the atmospheric composition of cooler gas giants.}


  \maketitle\ 

%

  \section{Introduction}
%

  Transiting long-period planets are valuable objects, but they are also challenging to characterize.
  These planets have a low transit probability, and giant planets have a lower occurrence rate than lower-mass planets (e.g., \citealt{mayor_2011,petigura_2018a,fulton_2021}),  which explains why so few of them have been detected so far. Besides, searching for the transit of long-period planets known from radial velocity surveys would be very time consuming, due to the low probability of transit.
  The four-year baseline of the Kepler mission \citep{borucki_2010} enabled the characterization of several transiting planets with orbital periods ranging from 100 to 300 days (e.g., \citealt{mancini_2016,dubber_2019,dalba_2021}),  and dedicated searches revealed a few dozen transiting planet candidates with orbital periods between 100 days to several years \citep{wang_2015,uehara_2016,foreman-mackey_2016}.
  
  The Transiting Exoplanet Survey Satellite (TESS) \citep{ricker_2015a} scans the whole sky, with each TESS sector being observed for 27 days almost continuously.
  As several subsequent sectors may overlap, some parts of the sky are monitored for up to one year with few interruptions.
  Exoplanets with orbital periods longer than 27 days will only show one transit event in a TESS sector. 
  Targeting single transit events opens the door to the discovery of planets with longer orbital periods than the ones found by searching periodic events. Several pipelines have thus been developed to search for single transit events (e.g., \citealt{gill_2020,montalto_2020a}).
  However, the period of these single-transiting candidates is mainly unconstrained.
  The nature of the TESS survey leads to multiple observations of the same sectors but about two years apart,
  and consequently the second transit of the same candidate may be detected. The range of possible orbital periods for a duo-transit candidate then becomes reduced to a discrete set of periods. \cite{osborn_2022b} have presented a new code to estimate the probability of each period alias.
  Their classification allows one to prioritize the follow-up of a duo-transit candidate by scheduling observations of the highest-probable
  periods first. Despite a strong detection bias against planets at long orbital periods,
  the TESS mission and the efficient follow-up effort has led to an increase
  of well-characterized transiting long-period giants (e.g., \citealt{gill_2020a,hobson_2021,dalba_2022,ulmer-moll_2022a}) .

  The CHaracterising ExOPlanet Satellite (CHEOPS; \citealt{benz_2021}) is the first European Space Agency mission 
  dedicated to the characterization of known exoplanets around bright stars.
  With space-grade photometric precision, CHEOPS can be used to confirm small transiting planets at long periods
  (e.g., \citealt{bonfanti_2021,lacedelli_2022}).
  For the multiplanetary system $\rm \nu^{2}\,Lupi$, the high signal-to-noise ratio of the CHEOPS observations
  allowed for precise measurement of the radius of its three small planets, as shown by \citet{delrez_2021}.
  Notably, CHEOPS has also been used to recover the orbital periods of TESS duo-transit candidates
  through targeted observations of the possible period aliases (e.g., Garai et al., Osborn et al., Tuson et al., in review).

  Long-period planets are ideal objects of study to gain insight into the formation and evolution of planetary systems, as their properties are less affected by
  stellar irradiation and stellar tidal interactions than close-in objects.
  For transiting planets, their planetary radius is measured with transit photometry, and their planetary mass and eccentricity are often measured through radial velocity follow-up or transit-timing variations, in the case of compact multiplanet systems.
  These fundamental parameters govern the planetary mean density and can lead to hints regarding a planet's interior structure.

  The study of modestly irradiated giants by \cite{miller_2011a} and \cite{thorngren_2016b} led to the definition of a planetary mass-metallicity relation,
  putting the metal enhancement of the Solar System giants Jupiter and Saturn in an exoplanetary context. Increasing the sample of
  long-period transiting exoplanets is essential to refining this relationship and, more broadly, understanding
  their formation and evolution processes.

  This paper reports the discovery and characterization of a long-period Neptune-mass planet transiting \ticA.
  The photometric and spectroscopic observations are described in Section~\ref{observations}, along with the observation strategy.
  Section ~\ref{methods} details the determination of the stellar parameters and the system modeling.
  The results of the joint modeling and the interior structure model are presented in Section~\ref{results}
  and the conclusions in Section~\ref{conclusion}.

\section{Observations}
\label{observations}

The discovery photometry was obtained from the TESS mission (Section~\ref{sec:tess}), and follow-up photometric observations were performed with the CHEOPS satellite (Section~\ref{sec:cheops}). Ground-based spectroscopic observations were carried out with the high-resolution spectrographs CORALIE (Section~\ref{sec:coralie}) and High Accuracy Radial velocity Planet Searcher (HARPS) (Section~\ref{sec:harps}).

\subsection{TESS photometry}
\label{sec:tess}

The TESS satellite \citep{ricker_2015a} observed \ticA\  during its primary and its first extended mission (year 1 and year 3). Until September 2023 (TESS cycle 5), no further TESS observations are planned. \ticA\ was observed at a two-minute cadence in three TESS sectors: sector 4 (2018-10-18 to 2018-11-15), sector 30 (from 2020-09-22 to 2020-10-21), and sector 31 (2020-10-21 to 2020-11-19).

We discovered a planet candidate in the TESS data using our specialized duo-transit pipeline \citep{tuson_2022}.
This pipeline was created to search for TESS duo-transits suitable for CHEOPS follow-up.
The pipeline concatenates the TESS Pre-search Data Conditioned Simple Aperture Photometry (PDCSAP) light curves from the primary and extended mission,
detrends the light curve using a mean sliding window, and then runs a box least squares (BLS; \citealt{kovacs_2002}) transit search
on the detrended light curve, with parameters optimized for duo-transit detection.
\ticA\ returned a duo-transit candidate with the first transit at the very beginning of the second orbit of sector 4 and a second transit in sector 30.
We checked both transits for asteroid crossing and centroid shifts (indicating a background eclipsing binary) before deciding to
pursue follow-up observations. This duo-transit candidate was also independently discovered by the Planet Hunter TESS team
and was announced as a Community Tess Object of Interest in March 2021.

The light curves from the primary and extended mission were obtained through the Mikulski Archive for Space Telescopes\footnote{\href{https://mast.stsci.edu/portal/Mashup/Clients/Mast/Portal.html}{mast.stsci.edu}} and the data reduction was done at the Science Processing Operation Center \citep{jenkins_2016}. We used the PDCSAP fluxes and their corresponding errors for our analysis.
We generated target pixel files with \texttt{tpfplotter} \citep{aller_2020} and used them to check for the presence of contaminant sources, down to a magnitude difference of seven. The apertures that were used to extract the light curves in all three of the sectors are not contaminated by neighboring stars. We did not include a dilution factor while modeling the TESS light curves because of the absence of contaminating sources.

\begin{table*}
  \caption{CHEOPS photometric observations of \ticA.}
  \label{table:table_cheops}
  \begin{tabular}{l l l l l l l}
    \hline
    \hline
    \noalign{\smallskip}
    Visit            & Start time           & End time                & Duration    & Texp        & Alias         & File reference\\
                     & [UT]                 & [UT]                    & [hours]     & [s]         & [days]        &      \\
    \hline
    \noalign{\smallskip}
    1                & 2021-09-24 00:54:22  & 2021-09-24 05:10:29     & 4.27        & 60          & 34.1          & CH\_PR110048\_TG006901\_V0200\\  
    2                & 2021-10-10 22:27:20  & 2021-10-11 02:40:27     & 4.22        & 60          & 39.8 \& 59.7  & CH\_PR110048\_TG007201\_V0200\\  
    3                & 2021-10-11 03:09:20  & 2021-10-11 07:36:28     & 4.45        & 60          & 39.8 \& 59.7  & CH\_PR110048\_TG007801\_V0200\\  
    4                & 2021-10-26 17:44:21  & 2021-10-26 23:24:30     & 5.67        & 60          & 31.1          & CH\_PR110048\_TG006701\_V0200\\  
    5                & 2021-11-03 17:56:20  & 2021-11-03 23:34:30     & 5.64        & 60          & 47.73         & CH\_PR110048\_TG007501\_V0200 \\ 
    \hline
  \end{tabular}
  \tablefoot{The Alias column corresponds to the period aliases compatible with an expected transit at the date of observation. Texp is the exposure time.}
\end{table*}

\subsection{CHEOPS photometry}
\label{sec:cheops}

As a space-based observatory, CHEOPS \citep{benz_2021} is  designed to perform ultra-high precision photometry on bright targets, one star at a time. Science observations with the satellite started in April 2020 and are mainly dedicated to the follow-up of known transiting exoplanets (e.g., \citealt{lendl_2020b,bonfanti_2021,szabo_2021,barros_2022}). The CHEOPS telescope has a 30\,cm effective aperture with a 1024\,x\,1024-pixel, back illuminated Charge-Coupled Device detector \citep{deline_2020}. The CHEOPS bandpass covers a broad range of optical wavelengths, from 330 to 1100\,nm. As shown by \citet{lendl_2020b}, CHEOPS is able to reach a photometric precision of ten parts per million (ppm) per one hour of binning for a very bright star (Vmag = 6.6). For faint stars, CHEOPS also complies with its requirement, as it provided a photometric precision of 75 ppm in three hours of integration time for a star of Vmag 11.9 \citep{benz_2021}.  

\begin{figure}
  \includegraphics[width=0.95\hsize]{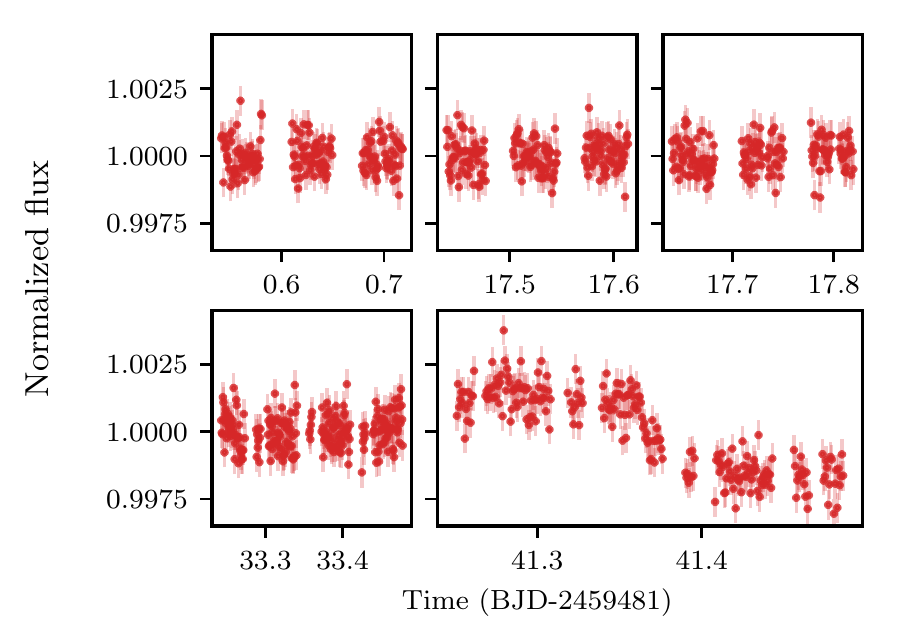}
  \caption{Raw CHEOPS light curve of \ticA\,b. The data covers five period aliases and displays an ingress in the bottom-right plot.}
  \label{fig:cheops_raw_lc}
\end{figure}

\ticA\ was observed with CHEOPS from 2021-09-24 to 2021-11-03 through the Duos program (CH\_PR110048: "Duos, Recovering long period duo-transiting planets"), which is part of the CHEOPS Guaranteed Time Observation program. This program aims to solve the orbital periods of TESS duo-transit candidates around stars with a V magnitude greater than 12. A detailed explanation of the Duos program and its scheduling strategy is presented in Section 3.2 of \citet{osborn_2022c}. The filler program targets candidates with a transit depth between two and four parts per thousand (ppt) such that the observation of a partial transit is sufficient to obtain a detection of the signal. Each CHEOPS visit lasts at least three CHEOPS orbits in order to correctly model instrumental trends and to identify the partial transit. Five CHEOPS visits of \ticA\ were observed with an exposure time of 60 seconds.
The details of the observations are presented in Table~\ref{table:table_cheops}.
The fifth visit of \ticA\ showed an ingress, and this transit event uniquely validated the orbital period of the candidate.
The CHEOPS light curve is shown in Figure~\ref{fig:cheops_raw_lc}, and no strong systematic trends are visible in the raw data.
After this visit, we stopped the monitoring of the target.

\ticA\ has a V magnitude of 11.4 and is one of the faintest targets to be observed with CHEOPS.
We used point spread function (PSF) photometry implemented in the PSF imagette photometric extraction package (PIPE\footnote{\href{https://github.com/alphapsa/PIPE}{github.com/alphapsa/PIPE}}) to reduce the CHEOPS data. In the case of \ticA, the PIPE data reduction was shown to perform better than the standard the data reduction pipeline version 13.1 (DRP\,13.1; \citep{hoyer_2020}). Over the five CHEOPS visits, the average mean absolute deviation of the PIPE light curve is 120 ppm lower than the DRP\,13.1 one using the default 25-pixel aperture. We chose to use the PIPE data reduction routines for our analysis. We selected data points without any recognized issue (flag equals zero). We removed data points for which the corresponding background flux has a value higher than 300\,$\rm e^{-}/pixel$. The background subtraction routine became unreliable due non-linear behavior beyond this threshold. Along with the time, flux, and errors on the flux, we extracted the background, roll angle ($\rm \theta$), and the X and Y coordinates to be used as detrending vectors in the analysis. We included detrending parameters, which improved the log-evidence values of the fit by a factor of five. The modulation in roll angle is correlated with the estimated contamination by nearby stars, but this contamination is quite low with a mean flux of 0.01\% compared to the target mean flux. We found that a linear combination of $\rm cos(2\theta)$ and $\rm cos(3\theta)$ was efficient at removing the roll angle dependence.
A table of the PIPE reduced CHEOPS photometric observations is available in a machine-readable format at the CDS.

\subsection{CORALIE spectroscopy}
\label{sec:coralie}

Spectroscopic vetting and radial velocity follow-up was carried out with CORALIE \citep{queloz_2001a}. The high-resolution fiber-fed spectrograph CORALIE is installed at the Nasmyth focus of the Swiss 1.2m Euler telescope (La Silla, Chile).\footnote{\href{https://www.eso.org/public/teles-instr/lasilla/swiss/}{eso.org/public/teles-instr/lasilla/swiss}}  CORALIE has a spectral resolution of 60,000, and the sampling per resolution element is equal to three pixels. Two fibers were used for the fiber injection. The first fiber was dedicated to the target observation, and the second fiber could collect light from a Fabry-Pérot etalon or the sky to allow for simultaneous wavelength calibration or background subtraction. 

\ticA\ was observed with CORALIE as part of a support program for CHEOPS to vet the candidates with spectroscopy and to evaluate the level of stellar activity visible in the radial velocity. Two high-resolution spectra were obtained several days apart. These observations are useful to rule out clear eclipsing binary scenarios. \ticA\ was then monitored at a low cadence to reveal possible long-term trends or strong activity signals.

We used the standard data reduction pipeline to extract the stellar radial velocities using the cross-correlation technique. The stellar spectrum of \ticA\ is cross-correlated with a G2 mask, which is the closest mask to the stellar type to obtain a cross-correlation function (CCF; e.g., \citealt{pepe_2002a}). The outputs of the pipeline are the radial velocity and its associated error as well as additional parameters derived from the CCF, such as the full width half maximum (FWHM), contrast, and bisector inverse slope (BIS).

We collected 14  radio velocity measurements of \ticA\ (from 2021-09-07 to 2022-01-23) with an exposure time varying between 1800 and 2100 seconds,  depending on the observing conditions. The radial velocity observations are detailed in Table \ref{table:table_rvs}.

\subsection{HARPS spectroscopy}
\label{sec:harps}

Once the orbital period of the candidate was solved with CHEOPS observations, we decided to add \ticA\ to an ongoing HARPS program dedicated to the follow-up of warm transiting giants (108.22L8.001, PI: Ulmer-Moll).
A fiber-fed spectrograph, HARPS has a spectral resolution of about 115,000. It is installed at the 3.6m telescope in La Silla, Chile \citep{mayor_2003a}.
\ticA\ was observed with the high-accuracy mode of HARPS from 2021-11-12 to 2022-02-15, and we obtained 23 HARPS measurements with an exposure time varying between 1800 and 2100 seconds. The average signal-to-noise ratio is equal to 40 at 550\,nm. The radial velocity observations are presented in Table \ref{table:table_rvs}.
The data reduction was performed with the standard data reduction pipeline \citep{lovis_2007a}.
The radial velocities were extracted with the cross-correlation technique using a G2 mask. Along with the radial velocities, the CCFs are used to derive several stellar activity indicators, such as the FWHM, contrast, and BIS. Additional parameters were also extracted from the spectra, namely, the S, $\rm H_{\alpha}$, Na, and Ca indexes. The HARPS spectra were co-added in order to derive stellar parameters, and the analysis is detailed in Section~\ref{stellar-analysis}.

\begin{table}
  \caption{Radial velocities of \ticA.}
  \label{table:table_rvs}
  \begin{tabular}{l l l l}
    \hline
    \hline
    \noalign{\smallskip}
    Time             & RV                   & RV error                  & Instrument\\
    BJD              & [$\rm ms^{-1}$]      & [$\rm ms^{-1}$]          & \\
    \hline
    \noalign{\smallskip}
    2459464.78269    & -3.86711             & 25.68928             & CORALIE\\
    2459470.72244    & 49.11076             & 27.98340             & CORALIE\\
    2459475.74732    & 36.17593             & 18.78204             & CORALIE\\
    ...&&&\\
    2459622.55655    & 55.81805             & 2.972752             & HARPS\\
    2459624.56960    & 51.25084             & 2.803944             & HARPS\\
    2459625.53665    & 54.48176             & 2.21336              & HARPS\\
    \hline
  \end{tabular}
  \tablefoot{Full table is available at CDS.}
\end{table}

\section{Methods}
\label{methods}

\subsection{Stellar parameter determination}
\label{stellar-analysis}

In order to derive precise stellar parameters, the HARPS spectra of \ticA\ were doppler-shifted according to their radial velocity measured during the data reduction and co-added in a single 1D master spectrum.
We used ARES+MOOG to derive stellar atmospheric parameters ($T_{\mathrm{eff}}$, $\log g$, microturbulence, [Fe/H]) following the same methodology described in \citet{santos_2013,sousa_2014}, and \citet{sousa_2021}. The latest version of ARES\footnote{The latest version, ARES v2, can be downloaded at https://github.com/sousasag/ARES.} \citep{sousa_2007,sousa_2015} was used to consistently measure the equivalent widths of selected iron lines based on the line list presented in \citet[][]{sousa_2008}. A minimization process was used in this spectral analysis to find ionization and excitation equilibrium and converge to the best set of spectroscopic parameters. The iron abundances were computed using a grid of Kurucz model atmospheres \citep{kurucz_1993} and the radiative transfer code MOOG \citep{sneden_1973}. The converged spectroscopic parameters are listed in Table \ref{table:stellar-params}.
The stellar abundances [Mg/H] = 0.07 $\pm$ 0.06 dex and [Si/H] = 0.00 $\pm$ 0.04 dex were derived using the classical curve-of-growth analysis method, assuming local thermodynamic equilibrium \citep[e.g.,][]{adibekyan_2012,adibekyan_2015}. The same codes and models were used for the abundance determinations. 

To determine the stellar radius of \ticA,\ we used a Markov chain Monte Carlo infrared flux method (MCMC IRFM; \citealt{blackwell_1977,schanche_2020}) that computes the stellar angular diameter and effective temperature via known relationships of these properties and the apparent bolometric flux. We used the stellar spectral parameters as priors to construct a spectral energy distribution from stellar atmospheric models that were attenuated via extinction estimates and used to compute the bolometric flux that is compared to the observed data taken from the most recent data releases for the bandpasses {\it Gaia} G, G$_{\rm BP}$, and G$_{\rm RP}$; 2MASS J, H, and K; and {\it WISE} W1 and W2 \citep{skrutskie_2006,wright_2010,gaiacollaboration_2021}, with the stellar atmospheric models taken from the \textsc{atlas} catalogs \citep{castelli_2003}. By converting the stellar angular diameter to the stellar radius using the offset corrected {\it Gaia} Early Data Release 3 parallax \citep{lindegren_2021}, we obtained $R_{\star}=0.938\pm0.007\, R_{\odot}$.

Finally, assuming $T_{\mathrm{eff}}$, [Fe/H], and $R_{\star}$ as input parameters, we determined the isochronal mass $M_{\star}$ and age $t_{\star}$ by using two different stellar evolutionary models. The first pair of mass and age values was computed through the isochrone placement algorithm \citep{bonfanti_2015,bonfanti_2016}, which interpolates the input parameters within pre-computed grids of PARSEC\footnote{\textsl{PA}dova and T\textit{R}ieste \textit{S}tellar \textit{E}volutionary \textit{C}ode: \url{http://stev.oapd.inaf.it/cgi-bin/cmd}.} v1.2S \citep{marigo_2017} isochrones and tracks. The second pair of mass and age values was retrieved by the Code Liègeois d'Èvolution Stellaire \citep[CLES;,][]{scuflaire_2008} code, which computes the best-fit stellar evolutionary tracks following the Levenberg-Marquadt minimization scheme as presented in \citet{salmon_2021}.
After carefully checking for the mutual consistency of the two respective pairs of outcomes through the $\chi^2$-based criterion broadly discussed in \citet{bonfanti_2021}, we merged the output distributions of both the stellar mass and age, ending up with the following estimates: $M_{\star}=0.905_{-0.041}^{+0.039}\,M_{\odot}$ and $t_{\star}=8.5\pm3.0$ Gyr.
The results are presented in Table~\ref{table:stellar-params}.

Following procedures going back to the CoRoT mission \citep{fridlund_2008},
we applied the Interactive Data Language package Spectroscopy Made Easy \citep{piskunov_1995, valenti_1996, piskunov_2017}.
With this code, and using the stellar parameters given above as fixed inputs, we then synthesized a spectrum based on
the well-determined stellar atmospheric grid MARCS 2012 \citep{gustafsson_2008} using atomic and molecular parameters
from the Vienna Atomic Line Database \citep{piskunov_1995} for the synthesis. Again following schemes outlined in \citet{fridlund_2020},
and references therein, and while keeping the turbulent velocities $\rm V_t$ macro and $\rm V_t$ micro fixed at the empirical values
found in the literature \citep[]{gray_2008, bruntt_2010, doyle_2014}, we found $\rm v \sin i$ to be 3.1$\pm$1 $\rm kms^{-1}$.

\begin{table}
  \caption{Stellar properties and stellar parameters.}
  \label{table:stellar-params}
  \centering
  \begin{tabular}{l c c}
    \hline
    \hline
    \noalign{\smallskip}
    & \ticA\  & \\
    \hline
    \noalign{\smallskip}
    Other Names & & \\
    \noalign{\smallskip}
    2MASS     & J03093220-3411530                        & 2MASS\\
    Gaia      & 5048310370810634624                      & Gaia\\
    TYC       & 7022-00871-1                             & Tycho\\
    TESS      & TIC\,209464063                           & TESS\\
    TOI       & \ticA                                    & TESS\\
    
    \hline
    \noalign{\smallskip}
    Astrometric Properties & & \\
    \noalign{\smallskip}
    R.A.                  & 03:09:32.2                       & TESS\\
    Dec                   & -34:11:54.53                     & TESS\\
    $\mu$R.A.(mas yr-1)   & -5.565±0.032                     & Gaia DR3\\
    $\mu$Dec.(mas yr-1)   & -89.15±0.04                      & Gaia DR3\\
    Parallax (mas)        & 6.099±0.014                       & Gaia DR3\\
    %
    \hline
    \noalign{\smallskip}
    Photometric Properties & & \\
    \noalign{\smallskip}
    TESS (mag)       & 10.669±0.006                           & TESS\\
    V (mag)          & 11.438±0.012                           & Tycho\\
    B (mag)          & 12.031±0.171                           & Tycho\\
    G (mag)          & 11.1483±0.0004                         & Gaia\\
    J (mag)          & 9.997±0.022                            & 2MASS\\
    H (mag)          & 9.699±0.021                            & 2MASS\\
    Ks(mag)          & 9.563±0.023                            & 2MASS\\
    W1 (mag)         & 9.533±0.023                            & WISE\\
    W2 (mag)         & 9.591±0.02                             & WISE\\
    W3 (mag)         & 9.552±0.033                            & WISE\\
    W4 (mag)         & 9.433±0.524                            & WISE\\
    TESS luminosity ($\rm L_{\odot}$)  & 0.740±0.025           & TESS\\
    \hline
    \noalign{\smallskip}
    Bulk Properties                        &                  & \\
    \noalign{\smallskip}
    $\rm T_{eff}$ (K)                       & $5485\pm63$      & Sec.~\ref{stellar-analysis} \\
    Spectral type                          & G7V              & Sec.~\ref{stellar-analysis}\\
    log g ($\rm cms^{-2}$)                 & $4.34\pm0.11$    & Sec.~\ref{stellar-analysis} \\
    $\rm V_t$ micro ($\rm kms^{-1}$)       & $0.82\pm0.03$    & Sec.~\ref{stellar-analysis} \\
    $\rm [Fe/H]$ (dex)                     & $0.00\pm0.01$          & Sec.~\ref{stellar-analysis} \\
    $\rm [Mg/H]$ (dex)                     & $0.07\pm0.06$          & Sec.~\ref{stellar-analysis} \\
    $\rm [Si/H]$ (dex)                     & $0.00\pm0.04$          & Sec.~\ref{stellar-analysis} \\
    $\rm v.sini$ ($\rm kms^{-1}$)          & $3.1\pm1.0$                  & Sec.~\ref{stellar-analysis} \\
    Age (Gyr)                           & $8.5\pm3.0$        & Sec.~\ref{stellar-analysis} \\
    Radius ($R_\odot$)                  & $0.938\pm0.007$     & Sec.~\ref{stellar-analysis} \\
    Mass ($M_\odot$)                    & $0.905_{-0.041}^{+0.039}$ & Sec.~\ref{stellar-analysis} \\

    \hline
  \end{tabular}
  \tablebib{2MASS \cite{skrutskie_2006}; GAIA EDR3 \cite{gaiacollaboration_2021}; TESS \citep{stassun_2019}; Tycho \citep{hog_2000}; WISE \cite{wright_2010}.}
\end{table}

\subsection{TESS photometric analysis}

\begin{figure}
  \includegraphics[width=0.95\hsize]{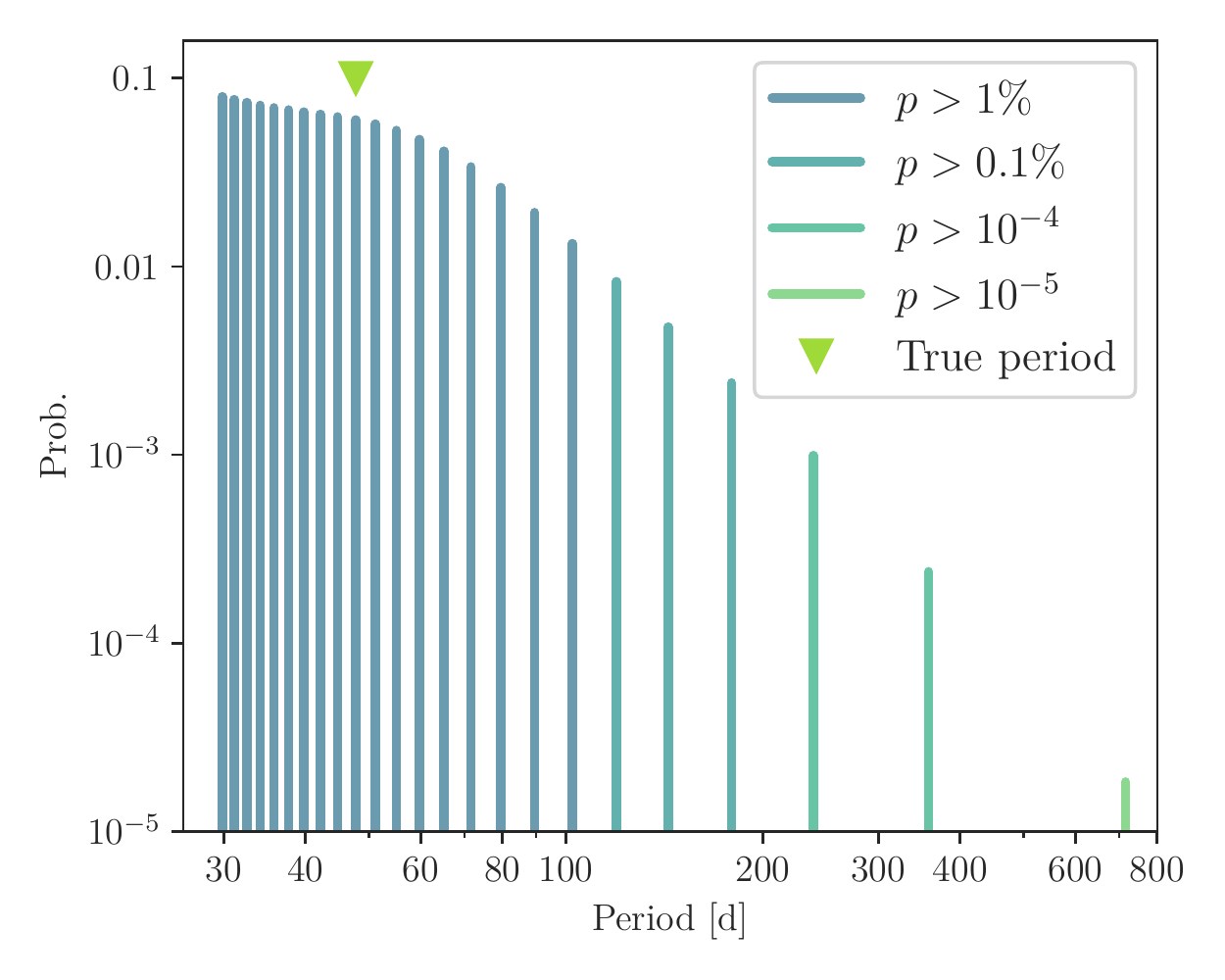}
  \caption{Marginalized probabilities of the possible period aliases, obtained with \texttt{MonoTools}, from TESS data only.
    The triangle indicates the true period of \ticA\,b.}
  \label{fig:period_aliases}
\end{figure}

We selected \ticA\ as a potential target for the CHEOPS Duos program, as its TESS light curve presents two single transits separated by a data gap of about 716 days. As for all targets in this CHEOPS program, we ran a photometric analysis with the software package \texttt{MonoTools} \citep{osborn_2022b}.\footnote{\href{https://github.com/hposborn/MonoTools}{github.com/hposborn/MonoTools}} 
\texttt{MonoTools} is especially adapted to detecting, vetting, and, modeling multiplanet systems in the case of single- and duo-transiting planets. The first step was to verify that the transit is not due to a rise in the background flux, which is indicative of an asteroid crossing or an artifact. We also tested that the transit model is a better fit to the event compared to a sine wave or a polynomial function. We checked that there was no feature in the X and Y coordinates coincident with the transit event, which would indicate a possible eclipsing binary or blend scenario. Finally, we confirmed that the shape, depth, and duration of the two transit events are compatible.

The second step was to model the duo-transit in order to determine which period aliases to observe.
\texttt{MonoTools} uses a Bayesian framework to marginalize over the possible period aliases and
compute the probability distribution of the period aliases, as shown in Figure~\ref{fig:period_aliases}.
This method fits the impact parameter, transit duration, and radius ratio in a way that is agnostic of the exoplanet period.
When combined with stellar parameters, such as bulk density \citep{stassun_2019}, the transit model allowed us to derive
an instantaneous transverse planetary velocity and, therefore, to compute a marginalized probability distribution for each allowed period alias.
This method uses a combination of the model likelihood and priors from a combination of window function and occurrence rate \citep[$P^{-2}$][]{kipping_2018},
a geometric transit probability ($p_g \propto a^{-1} \sim P^{-2/3}$; e.g., \citealt{winn_2010a}),
and an eccentricity prior \citep{kipping_2013a}.
For each allowed period alias, the eccentricity prior was applied using the eccentricity distribution
implied by the ratio of the transverse planetary velocity to the circular velocity at that period.

Once the probabilities of the different period aliases were computed, we organized the CHEOPS observations.
We scheduled 13 of the highest-probability aliases (covering a sum of 85\% of the probability space) on CHEOPS;
however, due to scheduling constraints, only a fraction of these are likely to be observed.
The transits of \ticA\,b are sufficiently deep such that CHEOPS does not require full transit coverage in order to successfully recover a transit.
We chose to only schedule partial transits, which required 2.8-orbit (4.6-hour) observations for each alias. We left a large degree of flexibility regarding the start time such that at least one orbit of in-transit data would always be observed.
This flexibility improves the likelihood of observations of \ticA\,b aliases being selected by the CHEOPS scheduler, even at low internal priority.

We realized a posteriori that the data from the first observation, labeled "visit 1" in Table~\ref{table:table_cheops}, covers only in-transit times. At first, this visit was inconclusive since we did not have any baseline signal. Further observations proved that visit 1 is in agreement with the baseline flux and rules out the 34.1-day period alias.
Visits 2 and 3 cover the period aliases 39.8\,days and 59.7\,days, respectively, and we obtained a clear non-detection of the transit signal.\footnote{One of the two visits would have been enough to exclude both aliases, but the Mission Planning System automatically scheduled both visits and we did not remove one of them.}
Visit 4 covers the 31.1-day period alias and also results in a clear non-detection.
Visit 5 covers the 47.73-day alias, and a transit ingress was detected.
We stopped the CHEOPS observations after the successful detection in the fifth visit.

\begin{figure}
  \includegraphics[width=\hsize]{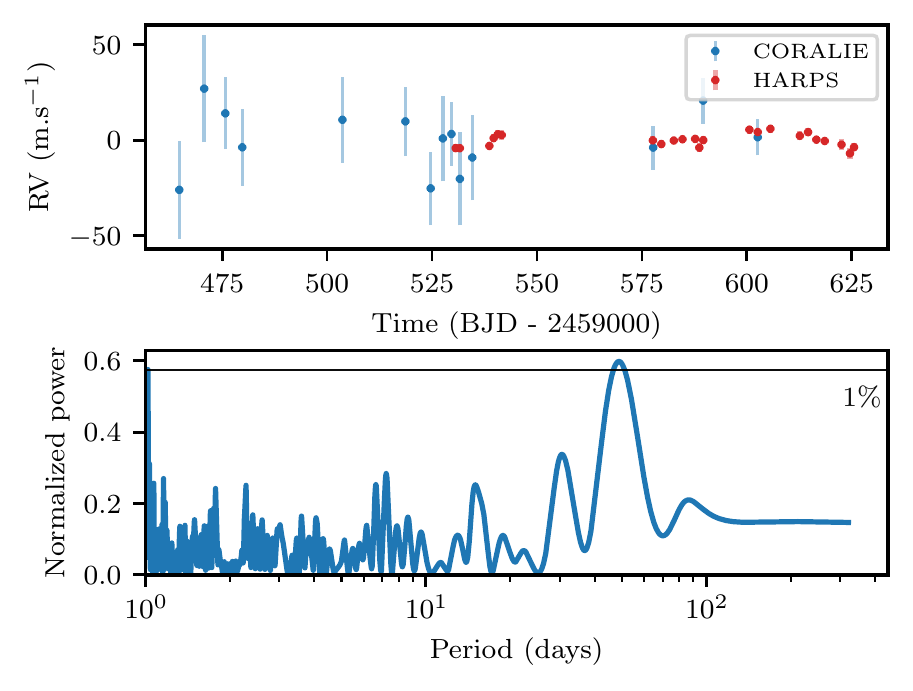}
  \caption{Top: Time series of radial velocities from CORALIE (blue dots) and HARPS (red triangles) for \ticA. The HARPS error bars are smaller than the marker. Bottom: Generalized Lomb-Scargle periodogram of the radial velocities (blue line). The 1\% false alarm probability level is indicated with a black line. The highest peak corresponds to a period of about 48.8 days.}
  \label{fig:rv_periodo_1240}
\end{figure}

\begin{figure}
  \includegraphics[width=\hsize]{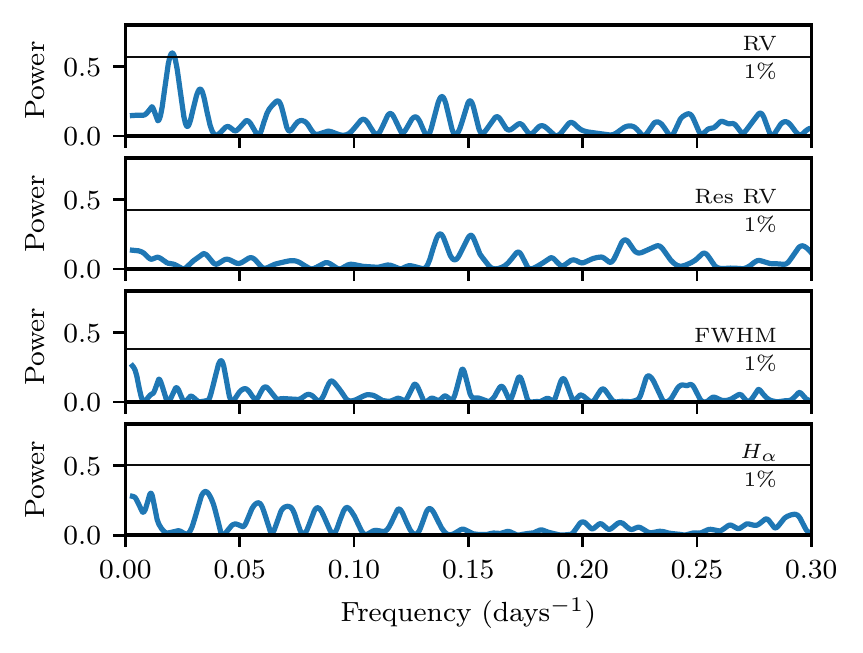}
  \caption{Generalized Lomb-Scargle periodogram of the radial velocities, radial velocity residuals, FWHM, and $H_{\alpha}$ index (from top to bottom).
    The 1\% false alarm probability level is indicated with a black line.}
  \label{fig:idx_periodo}
\end{figure}

\subsection{Radial velocity analysis}
\label{sec:rv}

The CORALIE observations of \ticA\ show a low radial velocity root mean square of $\rm 15\,ms^{-1}$
over two months, for an average radial velocity precision of  $\rm 20\,ms^{-1}$.
The preliminary analysis of the TESS data and the period recovery work done with CHEOPS result in a planetary candidate
with an orbital period of 47.73\,days and a radius of 4.9\,$\rm R_{\oplus}$. From these parameters, 
we estimated the expected mass of the planetary candidate to be between 18 and 21.5\,$\rm M_{\oplus}$
(based on the empirical mass-radius relations from \citealt{bashi_2017a} and \citealt{otegi_2020}, respectively),
leading to an expected semi-amplitude, ranging from 3.3 to $\rm 3.8\,ms^{-1}$.
With the HARPS Exposure Time Calculator,\footnote{\href{https://www.eso.org/observing/etc/bin/gen/form?INS.NAME=HARPS+INS.MODE=spectro}{eso.org/observing/etc}} we computed that the expected radial velocity precision
achievable with a 1800-second HARPS exposure is $\rm 1.4\,ms^{-1}$ for \ticA.
A detection of the radial velocity signal was feasible during the next few months of observability (November to February).
The target was then selected to be added to an ongoing HARPS program with the planned goal of characterizing long-period transiting exoplanets.
Figure~\ref{fig:rv_periodo_1240} presents the radial velocity dataset of \ticA\
and its corresponding generalized Lomb-Scargle (GLS) periodogram (\citealt{zechmeister_2009}), which combines both CORALIE and HARPS datasets.
No significant peak was found in the GLS periodogram of the following stellar activity indicators: FWHM and bisector
of the CCF or the $H_{\alpha}$ index. The GLS periodogram of the CCF contrast shows some peaks at periods unrelated to
the period of the planetary candidate.
Figure~\ref{fig:idx_periodo} displays the GLS periodogram of the CORALIE and HARPS radial velocities, the FWHM of the CCF, and the $H_{\alpha}$ index. The radial velocity residuals were calculated by removing the best-fit Keplerian model from the joint analysis in Section~\ref{sec:joint}.

A preliminary analysis of the radial velocity dataset was performed with the software package \texttt{kima} \citep{faria_2018a}, which is dedicated to the analysis of multiplanetary systems
and takes into account the radial velocity time series and their associated activity indicators.
Using a Bayesian framework, \texttt{kima} models radial velocity data as a sum of Keplerian functions.
Data from different spectrographs can be considered with a free radial velocity offset between datasets.
The number of planets is a free parameter of the model, and we chose to set it to a prior uniformly distributed as either zero, one, or two.
We set the maximum value of the orbital period prior to the time span of the dataset, and the upper bound of
the semi-amplitude prior was set to twice the peak-to-peak of the radial velocities.
Table~\ref{table:prior-kima} lists all priors used in this analysis.

The analysis of the CORALIE and HARPS data of \ticA\ with \texttt{kima} led to a detection
of one planetary signal with an orbital period of $48.63^{+2.61}_{-2.29}$\,days and a semi-amplitude of $\rm 3.94^{+0.79}_{-0.82}\,ms^{-1}$.
The Bayes factor between the no planet and one planet model is equal to 52 and 0.6 between the one planet and two planet model.
While the Bayes factor for the one planet model is not superior to 150, a common threshold for significance,
it indicates moderate evidence for this model \citep{feroz_2011}. The presence of transits with a compatible period adds credibility to the signal.

\begin{figure}
  \includegraphics[width=\hsize]{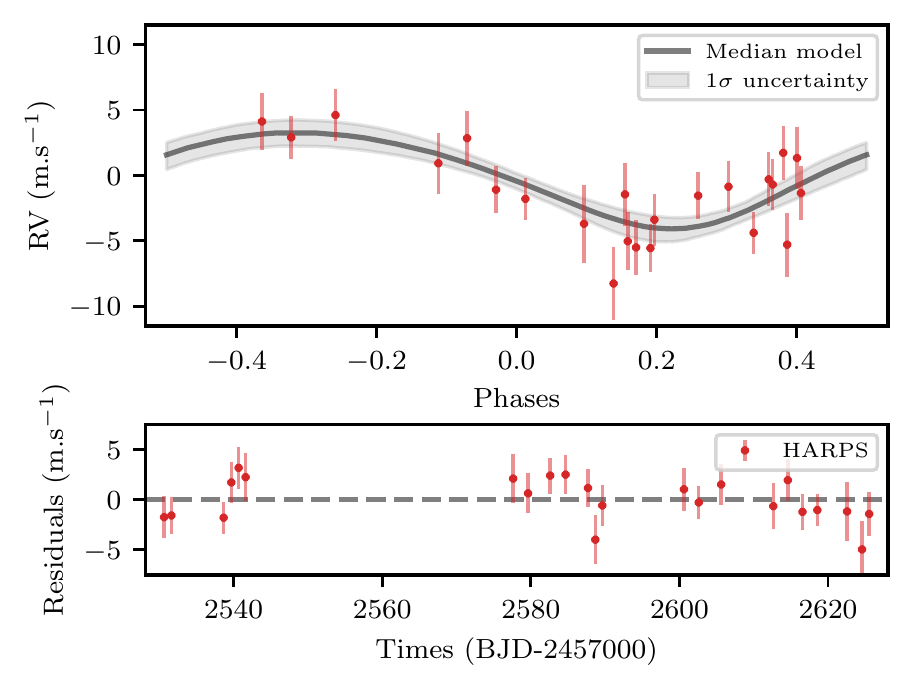}
  \caption{Radial velocities of \ticA. Top: HARPS radial velocities  (red dots) as a function of orbital phase. The median Keplerian model (gray line) is plotted along with its corresponding 1 $\sigma$ uncertainty (gray shaded area). Bottom: Radial velocity residuals (red dots) as a function of time.}
  \label{fig:rv_timeserie_1240}
\end{figure}

\begin{figure*}
  \includegraphics[width=0.95\hsize]{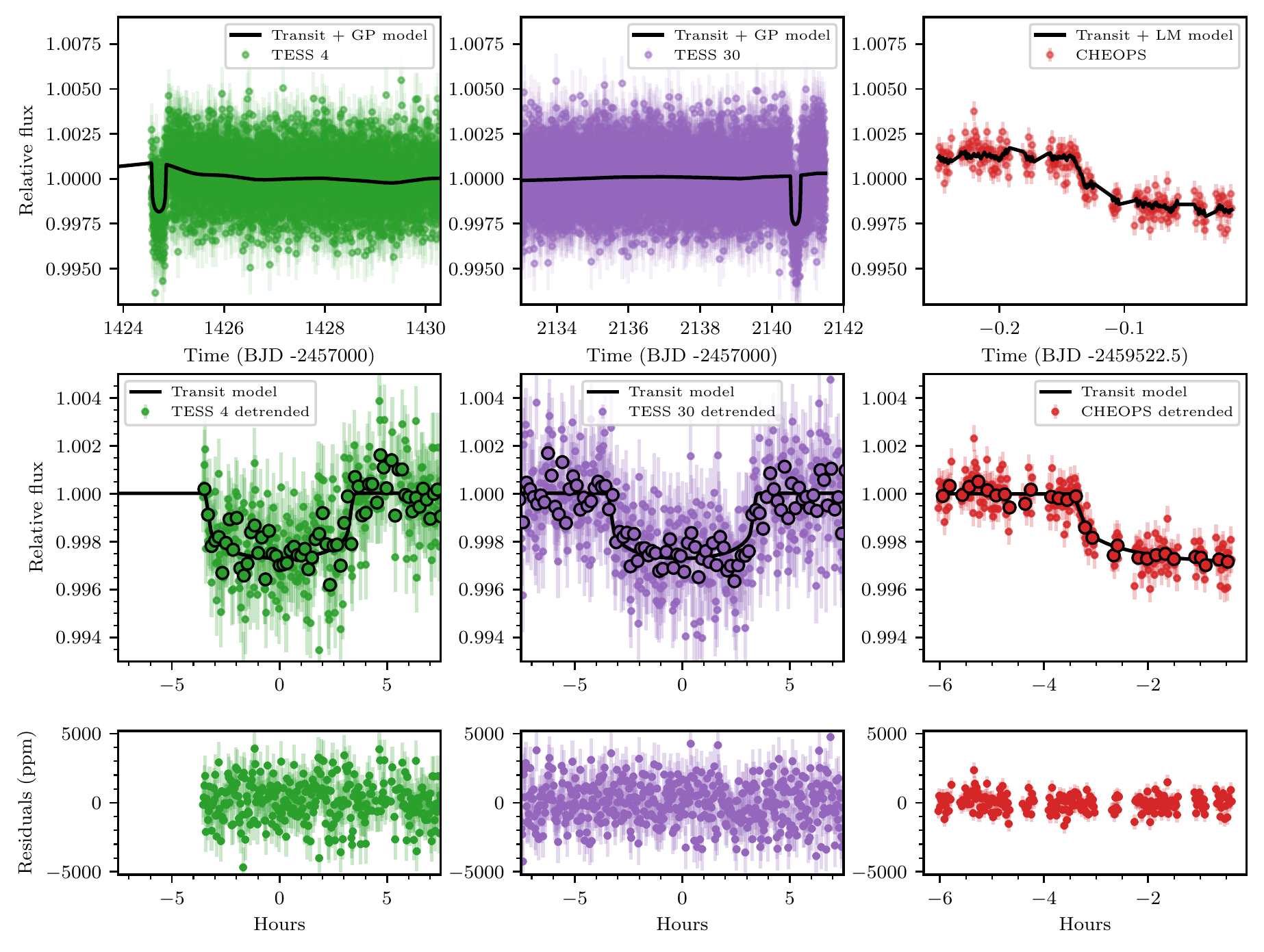}
  \caption{Photometric observations of \ticA. Top row: Light curves from both TESS sectors (green and purple dots) and CHEOPS observation (red dots). The full median model is plotted as a black line. Middle row: Detrended and phase-folded data from both TESS sectors (green and purple dots) and the CHEOPS observation (red dots). The phase-folded transit model is displayed in black. The dots outlined in black in all three panels show the ten-minute binned data. Bottom row: Residuals in parts per million between the full model and the respective light curve.}
  \label{fig:phase-folded_lc}
\end{figure*}

\subsection{Joint analysis}
\label{sec:joint}

We combined the photometric and radial velocity data to perform an analysis of the planetary system around \ticA.
We used the software package \texttt{juliet} \citep{espinoza_2019a} to jointly model the datasets, as 
\texttt{juliet} models multiplanetary systems, including transiting and non-transiting planets, within a Bayesian framework.
The light curve modeling in \texttt{juliet} is based on \texttt{batman} \citep{kreidberg_2015} for the model of the planetary transits.
The stellar activity and instrumental correlations can be specified with Gaussian processes or parametric functions, including linear models (LMs).
Photometric datasets from different instruments can be included in the model by adding independent
jitter and offset terms as well as dilution factors.
To model the Keplerian orbits in the radial velocity data, we used \texttt{RadVel} \citep{fulton_2018a}.
Several sampling methods can be implemented in \texttt{juliet}, and we chose the nested sampling
method \texttt{dynesty} \citep{speagle_2019}. Data from multiple instruments can be incorporated using separate values for
radial velocity jitter by allowing for offsets between the datasets.

Two planetary transits are visible in the TESS data in sectors 4 and 30. We chose to include these two relevant sectors
and to exclude sector 31, which does not contain any transit feature. Since only the fifth CHEOPS visit displays a partial transit of \ticA\,b,
we chose to only include this last visit for the joint modeling.
The model parameters for the planet are the orbital period, the planet-to-star radius ratio, the mid-transit time of the first transit,
the impact parameter, the argument of periastron, and the eccentricity.
Notably, we used a Beta prior for the eccentricity and a uniform prior between 40 and 60 days for the orbital period.
We compared the evidence values of different fits in order to choose the detrending of the TESS data (inclusion of Gaussian processes) and the modeling of the eccentricity.
We checked the impact of the eccentricity Beta prior by changing it from a Beta prior to a uniform prior between zero and one.
We found that both priors lead to similar posterior distributions as well as an eccentricity of  $\rm 0.14^{+0.07}_{-0.07}$ for the Beta
prior and $\rm 0.15^{+0.06}_{-0.08}$ for the uniform prior. Regarding the stellar density, we found that the density derived from the TESS photometric fit
with \texttt{MonoTools} is in agreement with the one derived in Section~\ref{stellar-analysis}.
In this joint fit, the stellar density is governed by a normal prior informed
by the stellar spectroscopic analysis done in Section~\ref{stellar-analysis}.
Because the TESS and CHEOPS bandpasses are different, we modeled the limb darkening
with a quadratic law using two different sets of q1 and q2 parameters \citep{kipping_2013}.
The priors used in the limb-darkening parameters were derived from the \texttt{LDCU}\footnote{\url{https://github.com/delinea/LDCU}} code, which is a modified version of the python routine implemented by \citet{espinoza_2015}.
It computes the limb-darkening coefficients and their corresponding uncertainties using
a set of stellar intensity profiles that account for the uncertainties in the stellar parameters.
The stellar intensity profiles were generated based on two libraries of synthetic stellar spectra:
ATLAS \citep{kurucz_1979} and PHOENIX \citep{husser_2013a}. Table~\ref{table:prior-juliet} details the priors used for all the model parameters.

The TESS photometric variability was taken into account in one Gaussian process for each sector using an approximate Matern kernel.
The CHEOPS light curves are known to have trends correlated with the roll angle of the spacecraft.
Therefore, we used LMs to decorrelate the CHEOPS data. The following parameters were scaled and then used
for linear decorrelation: background flux, cosine of the roll angle, and the X and Y coordinates.
A preliminary analysis of the CHEOPS data allowed us to select the relevant detrending parameters by comparing evidence values of the fits.
The functional form of the detrending for the fifth CHEOPS visit is presented in Section~\ref{sec:cheops} and in Table~\ref{table:system-parameters}. 
The priors on the associated coefficients are all uniform between -1 and one because the parameters were normalized and scaled. The photometric and instrumental variabilities were fitted simultaneously with the transit model.

The additional parameters included for the radial velocity data are the radial velocity semi-amplitude and jitters and offsets
for the two instruments (CORALIE and HARPS). The radial velocity analysis in Section~\ref{sec:rv} shows that the HARPS rms
is equal to $\rm 2\,ms^{-1}$. This value is comparable to the average uncertainty of the HARPS radial velocity ($\rm 1.4\,ms^{-1}$);
thus, we did not add parametric functions or Gaussian processes to model additional non-white noise.
We ran the final fit using the nested sampling algorithm with 2000 live points and until the uncertainty regarding the log evidence was smaller than 0.1.

\section{Results and discussion}
\label{results}

\subsection{Planetary system around TOI-5678}

Our joint analysis shows that \ticA\ hosts a transiting planet with an orbital period of 47.73\,days,
a planetary mass of $\rm 20\pm4\,M_{\oplus}$, and a planetary radius of $\rm 4.91\pm0.08\,R_{\oplus}$.
The transit has a total duration of $\rm 6.96\pm0.10\,hours,$ and the semi-major axis is $\rm 0.249\pm0.005\,au$.
The orbital eccentricity is equal to $\rm 0.14\pm0.07$ and is consistent at 2 \,$\rm \sigma$ with a circular orbit.
We note that a circular fit is only marginally disfavored when compared to a fit with free eccentricity ($\Delta\,ln\,Z\, <\, 2$).
We estimated the tidal circularization timescale with two formulas: one from \citet{adams_2006} and the other from \citet{jackson_2008a}.
We chose a tidal quality factor for the planet $\rm Q_{P}$ of $\rm 10^5$, which is within the range of values estimated
for the ice and gas giants of the Solar System \citep{banfield_1992,lainey_2009}.
The stellar tidal quality factor varied between $\rm 10^6$ and $\rm 10^8$ \citep{bonomo_2016}, without a significant effect on the timescale.
With both formalisms, we found a circularization timescale on the order of 20,000 to 25,000 Gyr.
This timescale is much larger than the age of the system ($\rm 8.5 \pm 3.0$ Gyr) and suggests that the orbit of
\ticA\,b is not subject to tidal circularization.

The HARPS phase-folded radial velocities are shown in Figure~\ref{fig:rv_timeserie_1240} with the median radial velocity model.
\ticA\,b induces a radial velocity signal with a semi-amplitude of $\rm 3.8\pm0.7\,ms^{-1}$. The HARPS jitter is $\rm 1.3 ^{+0.4} _{-0.2}\,ms^{-1}$
and the HARPS rms is about $\rm 2.1\,ms^{-1}$.
The CORALIE radial velocities were used for the joint fit; they are not displayed in Figure~\ref{fig:rv_timeserie_1240}
due to their large scatter and for improved readability.
After five months of radial velocity monitoring, we did not measure any significant drift.

Figure~\ref{fig:phase-folded_lc} presents the TESS light curves modeled with Gaussian process regression and the CHEOPS light curve modeled
with a combination of linear functions. The standard deviation of the two-minute, ten-minute, and one-hour binned residuals for the TESS light curves
are 1430, 650, 260\,ppm for sector 4 and 1480, 660, 265\,ppm for sector 30, respectively.
The full model of sector 4 appears flat in the top-left panel of Figure~\ref{fig:phase-folded_lc} because the variations of the Gaussian process
model are smaller in comparison to the transit depth.
The CHEOPS data has a standard deviation of the two-minute, ten-minute, and one-hour binned residuals of 450, 200, 110\,ppm.
The posterior distributions of the fitted parameters for \ticA\,b along with the stellar density and radial velocity semi-amplitude are presented in Figure~\ref{fig:corner_1240}.
Table~\ref{table:system-parameters} details the final parameters and associated uncertainties of \ticA\,b.

\subsection{TOI-5678\,b within the exoplanet population}

\ticA\,b joins the growing sample of transiting exoplanets with orbital periods larger than 15 days
and sizes between Neptune and Saturn (e.g., \citealt{dalba_2020,konig_2022}).
At the lower limit in mass of the group of planets between the ice and gas giants, \ticA\,b is an interesting object of study for its formation mechanism.
Under the core accretion scenario, the planetary core is formed early enough
to accrete gas from the protoplanetary disk, but a relatively small final mass indicates that gas accretion
has stopped and prevented the formation of a gas giant. 
For example, \cite{dransfield_2022} discovered a multiplanetary system of three long-period transiting sub-Neptunes,
with orbital periods of 22, 56, and 84\,days. The planets have very different mean densities,
which displays the different outcomes of planetary formation and evolution within one system.
The diversity of intermediate-mass exoplanets seems to be easily explained when studying the formation of Uranus and Neptune \citep{helled_2014,valletta_2022}.
The authors demonstrate that slight changes in the protoplanetary
disk and the environment of the planetary embryos (e.g., core accretion rate, solid surface density) lead to planets with greater differences in mass and composition.

\begin{figure}
  \includegraphics[width=\hsize]{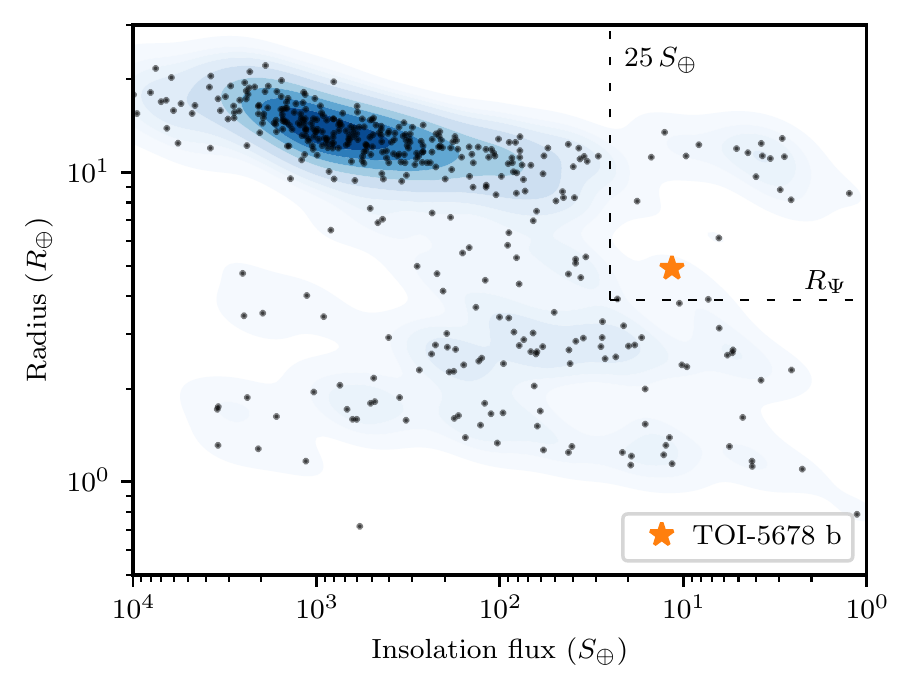}
  \caption{Planetary radius as a function of stellar insolation for exoplanets with radius and mass uncertainties lower than 8\% and 20\%, respectively.}
  \label{fig:radius_insolation}
\end{figure}

We present \ticA\,b within the population of transiting exoplanets in Figure~\ref{fig:radius_insolation}.
We downloaded the NASA exoplanet database\footnote{\href{https://exoplanetarchive.ipac.caltech.edu/cgi-bin/TblView/nph-tblView?app=ExoTbls&config=PS}{exoplanetarchive.ipac.caltech.edu}} on February 28, 2023, and selected confirmed exoplanets with radius uncertainties lower than 8\% and mass uncertainties lower than 20\%.
With a planetary radius of $\rm 4.91\pm0.08\,R_{\oplus}$ and a stellar insolation of $\rm 11.5\pm0.7\,S_{\oplus}$,
\ticA\,b is larger than the planets in the group of volatile-rich sub-Neptunes.
The planetary mean density of \ticA\,b is equal to $\rm 0.170\pm0.034\,\rho_{\oplus}$,
suggesting that the planet could host a large fraction of water or gas.
When compared to Neptune, \ticA\,b has a similar mass ($\rm 1.17\pm0.23\,M_{\Neptune}$)
but a larger radius ($\rm 1.262\pm0.021\,R_{\Neptune}$),
implying a lower planetary mean density ($\rm 0.57\,\rho_{\Neptune}$). 
We investigate the interior structure in detail in the following section.

Additional information on the formation and evolution of warm exoplanets
can be obtained through study of their atmospheres and their spin-orbit angles.
We estimated the transmission spectrum metric (TSM; \citealt{kempton_2018}) of \ticA\,b to be 40. This TSM value is rather low
but still places \ticA\,b in the top five targets with radii superior or equal to Neptune receiving less than $\rm 25\,S_{\oplus}$ (upper-right corner of Figure~\ref{fig:radius_insolation}).

\begin{figure}
  \includegraphics[width=\hsize]{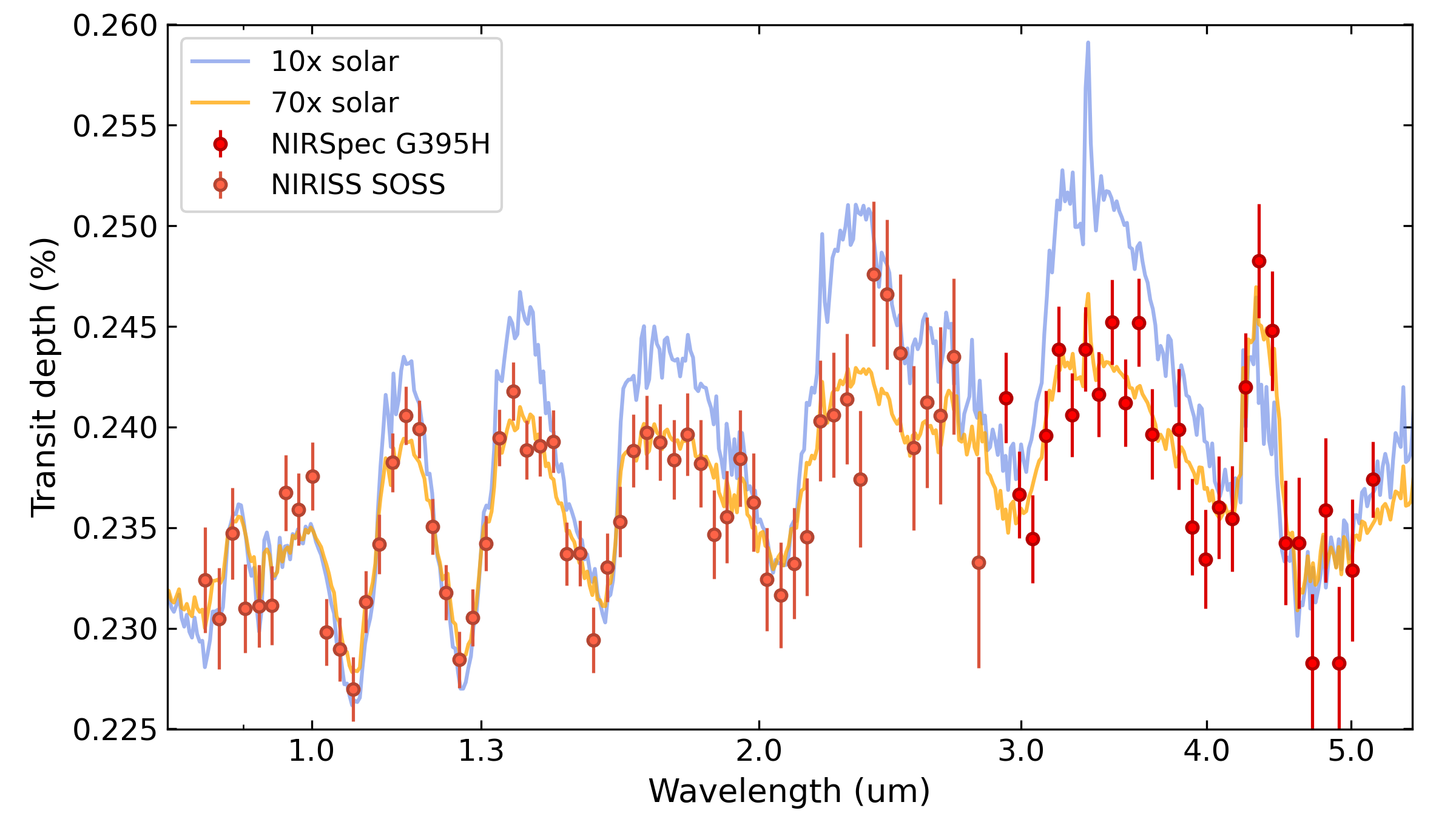}
  \caption{Modeled transmission spectra of the target in the NIRSpec/G395H and NIRISS wavelength ranges for two possible metallicities, 10x (blue line) and 70x (orange line) solar.
  Simulated flux measurements of each instrument are displayed as red points for NIRPSpec and as orange ones for NIRISS.}
  \label{fig:jwst}
\end{figure}

To illustrate the amplitude of the signal that would be seen by a James Webb Space Telescope (JWST) observation, we computed synthetic transmission spectra of TOI-5678~b
using the \textsc{Pyrat Bay} modeling framework \citep{cubillos_2021}. 
For simplicity, we assumed an atmosphere in thermochemical equilibrium at the planet's equilibrium temperature
and simulated two different metallicities: $70\times$ and $10\times$ solar.  For reference, the C/H ratio for Neptune and Uranus
(of similar mass to TOI-5678~b) is estimated to be $80 \pm 20\times$ solar \citep[e.g.,][]{atreya_2020}. Figure
\ref{fig:jwst} shows the model spectra along with simulated observations from the JWST NIRPSpec/G395H
and NIRISS/SOSS instruments \citep{batalha_2017}.
The spectra is generally dominated by H$_2$O bands, but a 3.5 $\mu$m
CH$_4$ feature and a 4.4 $\mu$m CO$_2$ feature are also visible.  The
two metallicity cases show detectable variations in the molecular
absorption features.

The planet is also suitable for obliquity measurement \citep{rossiter_1924,mclaughlin_1924}. We calculated that \ticA\,b produces a
Rossiter-McLaughlin effect of $\rm 7\,ms^{-1}$ (Eq 40 from \citealt{winn_2010a}),
a value large enough to be measured with state-of-the-art high-resolution spectrographs, such as ESPRESSO \citep{pepe_2021}.
Only a couple of gas giants with orbital periods longer than 40 days have an obliquity measurement, while no measurement has been done for
Neptune-type planets at such long periods. One the one hand, \citet{albrecht_2022} note that warm Jupiters and sub-Saturns with scaled semi-major axis ($\rm a / R_{\star}$)
larger than ten have relatively high obliquities. On the other hand, \citet{rice_2022} have found that tidally detached warm Jupiters appear more aligned than hot Jupiters;
however, the authors also show that Saturn-type objects can have a wide range of obliquities. An obliquity measurement for \ticA\ can help shed
light on its formation and evolution mechanisms.

\subsection{Formation and composition of TOI-5678\,b}
\label{internal_struct}

\begin{figure}
  \includegraphics[width=\hsize]{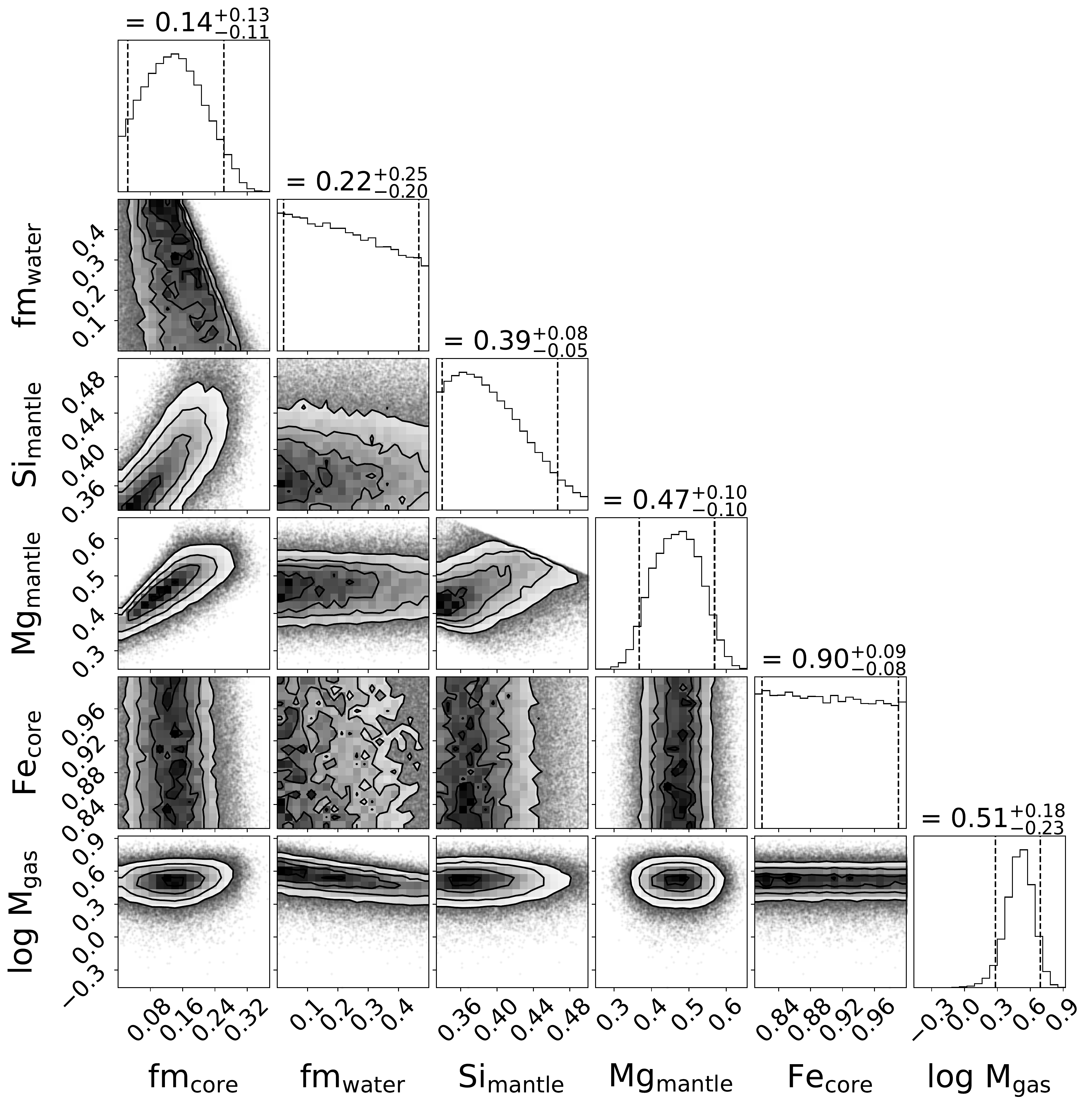}
  \caption{Posterior distributions of the internal structure parameters of TOI-5678\,b. The median and the five and 95 percentiles shown in the title of each column. The depicted internal structure parameters are: The layer mass fractions of the iron core and water layer with respect to the solid planet, the molar fractions of Si and Mg in the mantle and Fe in the core, and the logarithm of the gas mass in Earth masses.}
  \label{fig:int_structure}
\end{figure}

We applied a Bayesian inference scheme to model the internal structure of \ticA\,b.
A thorough description of the used method, which is based on \citet{dorn_2017b}, can be found in \citet{leleu_2021}.
In the following, we briefly summarize the most important aspects of the model:
the assumed priors, the input parameters, and the forward model that was used to calculate the likelihood of a set of internal structure parameters.

For the priors, we assumed a distribution that is uniform on the simplex for the layer mass fractions of the iron core,
silicate mantle, and the water layer, all with respect to the solid core of the planet (without the H/He atmosphere layer).
We set an upper limit of 50\% for the water mass fraction, following \citet{thiabaud_2014}. For the gas mass fraction,
a log-uniform distribution was assumed. It is important to note that the results of the internal structure model depend
to a certain extent on the chosen priors. The Bayesian inference model takes both planetary and stellar observables
as input parameters: the mass, radius, age, effective temperature,
and molecular abundances ([Si/H], [Mg/H], and [Fe/H]) of the star and the transit depth,
relative mass, and period of the planet. 

Finally, the forward model we selected assumes a spherically symmetric planet with four fully distinct layers
(iron core, silicate mantle, water, and H/He atmosphere). It uses equations of state from \citet{hakim_2018},
\citet{sotin_2007}, \citet{haldemann_2020}, and \citet{lopez_2014}.
We stress that in the current version of our model, the H/He atmosphere was modeled independently
from the solid part of the planet, and we assumed a fixed temperature and pressure at the water-gas boundary,
thereby neglecting the effects of the gas layer on the remaining layers of the planet.
Moreover, we made the assumption that the Si/Mg/Fe ratios of the planet
and the star are identical (see e.g., \citet{thiabaud_2015}).
We note that more recent results by \citet{adibekyan_2021} confirm
that there is a correlation between the composition of a star and its planets,
but the authors suggest a relationship that is not 1:1.

Figure~\ref{fig:int_structure} shows the results of the internal structure model.
While the mass of the H/He layer of TOI-5678\,b is reasonably well constrained
and quite large, with $3.2^{+1.7}_{-1.3}$\,M$_{\oplus}$, the posterior of the water mass fraction
is almost completely unconstrained.
The molar fraction of iron in the core ($\rm Fe_{core}$) is also unconstrained and only varies between the bounds allowed by its prior.
Furthermore, the posterior distribution of the radius
of the gas layer has a median of $2.53^{+0.32}_{-0.34}$\,R$_{\oplus}$.
By combining the planet mass with the inferred mass of H-He, we predicted a minimum mass of 11.1 M$_{\oplus}$
of heavy elements for TOI-5678 b. A planet with such a large content of heavy elements likely accreted
its core beyond the ice line (e.g., \citealt{venturini_2020a, venturini_2020}).

We compared TOI-5678\,b to synthetic planets with similar parameters produced with the New Generation Planetary Population Synthesis \citep{mordasini_2009}, the latest version of the Bern model \citep{emsenhuber_2021b,emsenhuber_2021a} available on DACE.\footnote{\href{https://dace.unige.ch/populationAnalysis/?populationId=NG76}{dace.unige.ch/populationAnalysis}} We chose the NG76 simulation where 100 embryos are used. We found that synthetic planets with a radius, mass, and semi-major axis close to TOI-5678b form at a couple of astronomical units from their host star. After a first initial growth up to about 10\,M$_{\oplus}$, the planets migrate inwards while steadily accreting mass but without reaching a mass high enough to trigger rapid gas accretion. These planets are part of the system architecture called migrated sub-Neptunes (Class II), as described in \citet{emsenhuber_2023}. We note that some planets can also reach about 20\,M$_{\oplus}$ by accreting a significant amount of their mass through collisions. These planets forming through giant impacts originate from inside the ice line, and their final planetary radius is usually below 3\,R$_{\oplus}$. In contrast, planets originating from outside the ice line have a larger radius, and they contain a significant fraction of gas, similar to what we estimated for TOI-5678\,b. The knowledge of both mass and radius of TOI-5678\,b allows us to favor a formation and migration scenario starting outside of the ice line.
This is also supported by other formation models, which find that only water-rich planets account for planets with masses larger than 10\,M$_{\oplus}$ and radii larger than 3\,R$_{\oplus}$ (\citealt{venturini_2020a},  Figs D1 and D2). This support reinforces the likelihood that TOI-5678\,b formed beyond the ice line and migrated inwards later on.
Finally, the nearly circular orbit of TOI-5678\,b and long circularization timescale suggest that the hypothesis of a planet-disk migration is favored over  a high eccentricity migration scenario. This migration hypothesis can be tested via the measurement of the spin-orbit alignment of the system.


\begin{table}
  \caption{Fitted and derived parameters for \ticA\,b.}
  \label{table:system-parameters}
  \centering
  \begin{tabular}{l c}
    \hline
    \hline
    \noalign{\smallskip}
    Parameters                                 & \ticA\,b             \\
    \hline
    \noalign{\smallskip}
    Fitted parameters                          &                   \\
    \noalign{\smallskip}
    Orbital period (days)                      & $47.73022 ^{+0.00014} _{-0.00013}$\\
    Time of transit T0 (days)                  & $2458424.7059 ^{+0.0035} _{-0.0032}$\\
    Radius ratio                               & $0.0480 ^{+0.0007} _{-0.0007}$ \\
    Impact parameter                           & $0.17 ^{+0.13} _{-0.10}$\\
    Stellar density ($\rm kg.m^{-3}$)           & $1551 ^{+83} _{-95}$          \\ 
    TESS limb darkening q1                     & $0.351 ^{+0.027} _{-0.028}$   \\
    TESS limb darkening q2                     & $0.302 ^{+0.010} _{-0.011}$   \\
    CHEOPS limb darkening q1                   & $0.46 ^{+0.04} _{-0.04}$      \\
    CHEOPS limb darkening q2                   & $0.363 ^{+0.013} _{-0.013}$  \\
    Eccentricity                               & $0.14 ^{+0.07} _{-0.07}$     \\
    Argument of periastron                     & $208 ^{+18} _{-11}$           \\
    Radial velocity semi-amplitude ($\rm ms^{-1}$)            & $3.8 ^{+0.7} _{-0.7}$   \\
    \hline
    \noalign{\smallskip}
    Derived parameters                         &                  \\
    \noalign{\smallskip}
    Planetary radius ($R_{\oplus}$)             &  $4.91 ^{+0.08} _{-0.08}$ \\
    Planetary mass ($M_{\oplus}$)               &  $20 ^{+4} _{-4}$  \\
    Inclination (degrees)                      &  $89.83 ^{+0.11} _{-0.12}$ \\
    Semi major axis (au)                       &  $0.249 ^{+0.005} _{-0.005}$ \\
    Transit duration (hours)                   &  $6.96 ^{+0.10} _{-0.10}$    \\
    Equilibrium temperature (K)                &  $513 ^{+8} _{-7}$    \\

    \hline
    \noalign{\smallskip}
    Instrumental parameters                    &                  \\
    \noalign{\smallskip}
    TESS offset                                & $-0.00002 ^{+0.00009} _{-0.00008}$  \\
    TESS jitter (ppm)                          & $1.9 ^{+13.6} _{-1.6}$                 \\
    TESS 2 offset                              & $-0.00002 ^{+0.00005} _{-0.00005}$   \\
    TESS 2 jitter (ppm)                        & $1.5 ^{+11.2} _{-1.3}$              \\
    CHEOPS offset                              & $-0.00121 ^{+0.00008} _{-0.00008}$    \\
    CHEOPS jitter (ppm)                        & $123 ^{+84} _{-91}$              \\

    \noalign{\smallskip}
    GP amplitude TESS  (relative flux)           & $0.000028 ^{+0.00009} _{-0.00006}$ \\
    GP time-scale TESS  (days)                   & $0.9 ^{+0.4} _{-0.3}$            \\
    GP amplitude TESS 2  (relative flux)         & $0.00015 ^{+0.00007} _{-0.00004}$  \\
    GP time-scale TESS 2  (days)                 & $1.2 ^{+0.09} _{-0.5}$           \\

    \noalign{\smallskip}
    Background CHEOPS (relative flux)           & $-0.00014 ^{+0.00006} _{-0.00006}$\\
    cos($\rm 2\theta$) CHEOPS (days)            & $0.00023 ^{+0.00011} _{-0.00012}$ \\
    cos($\rm 3\theta$) CHEOPS (days)            & $0.00014 ^{+0.00008} _{-0.00008}$\\ 
    X coordinate CHEOPS                         & $-0.00020 ^{+0.00005} _{-0.00005}$ \\
    Y coordinate CHEOPS                         & $-0.00016 ^{+0.00005} _{-0.00005}$\\

    \noalign{\smallskip}
    CORALIE offset ($\rm ms^{-1}$)            & $24.0^{+4}_{-4}$     \\
    HARPS offset ($\rm ms^{-1}$)              & $59.7^{+0.5}_{-0.5}$     \\ 
    CORALIE jitter ($\rm ms^{-1}$)            & $3.0^{+4}_{-2}$     \\
    HARPS jitter ($\rm ms^{-1}$)              & $1.3^{+0.4}_{-0.2}$   \\
    
  \end{tabular}
\end{table}


\section{Conclusions}
\label{conclusion}

This paper reports the discovery \ticA\,b, a transiting long-period Neptune-mass planet.
Its host is a G7V-type star of solar metallicity with an age of $8.5\pm3.0$ Gyrs.
The planet itself is on an almost circular orbit and has an orbital period of 47.73\,days.
Its planetary mass is equal to $\rm 20\pm4\,M_{\oplus},$ and its planetary radius is equal to $\rm 4.91\pm0.08\,R_{\oplus}$.
We used the space-based photometric satellites TESS and CHEOPS to identify and confirm the orbital period of \ticA\,b. 
We combined the photometric data with CORALIE and HARPS radial velocities to characterize the system and
measure the mass and eccentricity of the transiting companion. We modeled the interior structure of \ticA\,b
and constrained the mass of its H/He layer to $3.2^{+1.7}_{-1.3}$\,M$_{\oplus}$, assuming the four-layer model presented in Section~\ref{internal_struct}.
We show that TOI-5678\,b likely formed beyond the ice line and then migrated inwards to its current location.
The long orbital period of \ticA\,b prevents it from undergoing strong dynamical interactions with the host star. Additional observations, such as Rossiter-McLaughlin measurements, would further improve our knowledge of the system.
Such long-period intermediate-mass planets as \ticA\,b provide the opportunity to bridge exoplanet
and Solar System science, providing key information on the Uranus- and Neptune-like planets.

\subsection*{Acknowledgments}

This work has been carried out within the framework of the National Centre of Competence
in Research PlanetS supported by the Swiss National Science Foundation under grants 51NF40\_182901 and 51NF40\_205606.
The authors acknowledge the financial support of the SNSF.
AT acknowledges support from an STFC PhD studentship.
ML acknowledges support of the Swiss National Science Foundation under grant number PCEFP2194576. 
PM acknowledges support from STFC research grant number ST/M001040/1.
JE, YA, and MJH acknowledge the support of the Swiss National Fund under grant 200020\_172746.
ABr was supported by the SNSA.
MF and CMP gratefully acknowledge the support of the Swedish National Space Agency (DNR 65/19, 174/18, 177/19).
DG and LMS gratefully acknowledges financial support from the CRT foundation under Grant No. 2018.2323 ``Gaseous or rocky? Unveiling the nature of small worlds''.
CM acknowledges the support from the SNSF under grant 200021\_204847 ``PlanetsInTime’'.
S.G.S. acknowledge support from FCT through FCT contract nr. CEECIND/00826/2018 and POPH/FSE (EC).
ACC and TW acknowledge support from STFC consolidated grant numbers ST/R000824/1 and ST/V000861/1, and UKSA grant number ST/R003203/1.
We acknowledge support from the Spanish Ministry of Science and Innovation and the European Regional Development Fund through grants ESP2016-80435-C2-1-R, ESP2016-80435-C2-2-R, PGC2018-098153-B-C33, PGC2018-098153-B-C31, ESP2017-87676-C5-1-R, MDM-2017-0737 Unidad de Excelencia Maria de Maeztu-Centro de Astrobiologí­a (INTA-CSIC), as well as the support of the Generalitat de Catalunya/CERCA programme. The MOC activities have been supported by the ESA contract No. 4000124370.
S.C.C.B. acknowledges support from FCT through FCT contracts nr. IF/01312/2014/CP1215/CT0004.
XB, SC, DG, MF and JL acknowledge their role as ESA-appointed CHEOPS science team members.
This project was supported by the CNES.
The Belgian participation to CHEOPS has been supported by the Belgian Federal Science Policy Office (BELSPO) in the framework of the PRODEX Program, and by the University of Liège through an ARC grant for Concerted Research Actions financed by the Wallonia-Brussels Federation; L.D. is an F.R.S.-FNRS Postdoctoral Researcher.
This work was supported by FCT - Fundação para a Ciência e a Tecnologia through national funds and by FEDER through COMPETE2020 - Programa Operacional Competitividade e Internacionalizacão by these grants: UID/FIS/04434/2019, UIDB/04434/2020, UIDP/04434/2020, PTDC/FIS-AST/32113/2017 \& POCI-01-0145-FEDER- 032113, PTDC/FIS-AST/28953/2017 \& POCI-01-0145-FEDER-028953, PTDC/FIS-AST/28987/2017 \& POCI-01-0145-FEDER-028987, O.D.S.D. is supported in the form of work contract (DL 57/2016/CP1364/CT0004) funded by national funds through FCT.
This project has received funding from the European Research Council (ERC) under the European Union’s Horizon 2020 research and innovation programme (grant agreement SCORE No 851555).
J.H. supported by the Swiss National Science Foundation (SNSF) through the Ambizione grant \#PZ00P2\_180098.
P.E.C. is funded by the Austrian Science Fund (FWF) Erwin Schroedinger Fellowship, program J4595-N.
B.-O. D. acknowledges support from the Swiss State Secretariat for Education, Research and Innovation (SERI) under contract number MB22.00046
and from the Swiss National Science Foundation (PP00P2-190080).
This project has received funding from the European Research Council (ERC) under the European Union’s Horizon 2020 research and innovation programme (project {\sc Four Aces}; grant agreement No 724427). It has also been carried out in the frame of the National Centre for Competence in Research PlanetS supported by the Swiss National Science Foundation (SNSF). DE acknowledges financial support from the Swiss National Science Foundation for project 200021\_200726.
M.G. and V.V.G. are F.R.S.-FNRS Senior Research Associates.
SH gratefully acknowledges CNES funding through the grant 837319.
This work was granted access to the HPC resources of MesoPSL financed by the Region Ile de France and the project Equip@Meso (reference ANR-10-EQPX-29-01) of the programme Investissements d'Avenir supervised by the Agence Nationale pour la Recherche.
LBo, GBr, VNa, IPa, GPi, RRa, GSc, VSi, and TZi acknowledge support from CHEOPS ASI-INAF agreement n. 2019-29-HH.0.
This work was also partially supported by a grant from the Simons Foundation (PI Queloz, grant number 327127).
IRI acknowledges support from the Spanish Ministry of Science and Innovation and the European Regional Development Fund through grant PGC2018-098153-B- C33, as well as the support of the Generalitat de Catalunya/CERCA programme.
GyMSz acknowledges the support of the Hungarian National Research, Development and Innovation Office (NKFIH) grant K-125015, a a PRODEX Experiment Agreement No. 4000137122, the Lend\"ulet LP2018-7/2021 grant of the Hungarian Academy of Science and the support of the city of Szombathely.
NAW acknowledges UKSA grant ST/R004838/1.
KGI is the ESA CHEOPS Project Scientist and is responsible for the ESA CHEOPS Guest Observers Programme. She does not participate in, or contribute to, the definition of the Guaranteed Time Programme of the CHEOPS mission through which observations described in this paper have been taken, nor to any aspect of target selection for the programme.
The authors acknowledge the use of public TESS data from pipelines at the TESS Science Office and at the TESS Science Processing Operations Centre.

\bibliographystyle{aa}
\bibliography{Photometry-Transit}

%
%
\begin{appendix}

  \section{Radial velocity modeling priors}
  
  \begin{table*}[h!]
  \caption{Priors for the radial velocity fit.}
  \label{table:prior-kima}
  \centering
  \begin{tabular}{l l l}
    \hline
    \hline
    \noalign{\smallskip}
    Parameters                                 & Distribution     & Value \\
    \hline
    \noalign{\smallskip}
    Number of planets                          & Uniform          & (0, 1)   \\
    Orbital period (days)                      & LogUniform       & (1, 161)       \\
    Semi-amplitude ($\rm ms^{-1}$)             & LogUniform       & (1, 136)       \\ 
    Eccentricity                               & Kumaraswamy      & (0.867, 3.03)       \\
    Mean anomaly                               & Uniform          & (-$\pi$, $\pi$)       \\
    Argument of periastron                     & Uniform          & (0, 2$\pi$)       \\
    Jitter ($\rm ms^{-1}$)                     & LogUniform       & (1, 200)       \\
    \hline
  \end{tabular}
  \end{table*}

  \section{Joint modeling priors}
  \label{appendix:prior_juliet}

  \begin{table*}[h!]
  \caption{Priors for the joint modeling of photometric and radial velocity data.}
  \label{table:prior-juliet}
  \centering
  \begin{tabular}{l l c}
    \hline
    \hline
    \noalign{\smallskip}
    Parameters                                 & Distribution     & \ticA\,b\\
    \hline
    \noalign{\smallskip}
    Orbital period (days)                      & Uniform          & (40, 60)\\
    Time of transit T0 (days)                  & Uniform          & (2458485.9, 2458486.3)\\
    Radius ratio                               & Uniform          & (0, 1)\\
    Impact parameter                           & Uniform          & (0, 1) \\
    Stellar density                            & Normal           & (685.24, 79.18)\\ 
    TESS limb darkening q1                     & Normal           & (0.274, 0.027)\\
    TESS limb darkening q2                     & Normal           & (0.275, 0.029)\\
    CHEOPS limb darkening q1                   & Uniform          & (0.47, 0.05) \\
    CHEOPS limb darkening q2                   & Uniform          & (0.363, 0.015) \\
    Eccentricity                               & Kumaraswamy / fixed     & (0.867, 3.03)\\
    Argument of periastron                     & Uniform          & (0, 2$\pi$)\\
    TESS offsets                               & Normal           & (0, 0.01)\\
    TESS jitters (ppm)                         & LogUniform       & (0.1, 1000) \\
    CHEOPS offsets                             & Normal           & (0, 0.01) \\
    CHEOPS jitters (ppm)                       & LogUniform       & (0.1, 1000) \\

    \noalign{\smallskip}
    GP amplitude TESS (relative flux)           & LogUniform     & (1e-06, 100.0)\\
    GP time-scale TESS (days)                   & LogUniform     & (0.001, 100.0)\\
    GP amplitude TESS 2  (relative flux)        & LogUniform     & (1e-06, 100.0)\\
    GP time-scale TESS 2  (days)                & LogUniform     & (0.001, 100.0)\\
    
    \noalign{\smallskip}
    Background CHEOPS (relative flux)           & LogUniform     & (-1, 1) \\
    cos($\rm 2\theta$) CHEOPS (days)            & Uniform        & (-1, 1) \\
    cos($\rm 3\theta$) CHEOPS (days)            & Uniform        & (-1, 1) \\ 
    X coordinate CHEOPS                         & LogUniform     & (-1, 1) \\
    Y coordinate CHEOPS                         & Uniform        & (-1, 1)\\

    \noalign{\smallskip}
    Semi-amplitude ($\rm kms^{-1}$)             & Uniform         & (0, 100)\\
    Spectrograph offsets ($\rm kms^{-1}$)       & Uniform         & (-100, 100) \\ 
    Spectrograph jitters ($\rm kms^{-1}$)       & LogUniform      & (0.001, 0.2)\\

    \hline
  \end{tabular}
  \tablefoot{The normal distribution is defined by its mean and variance. The uniform and log-uniform distributions were defined with the lower and upper bounds of the distribution. The priors on the radial velocity offsets and jitters are identical for CORALIE and HARPS.}
  \end{table*}

  \section{Joint modeling posterior distribution}
  \label{appendix:corner_juliet}

  \begin{figure*}
    \includegraphics[width=0.95\hsize]{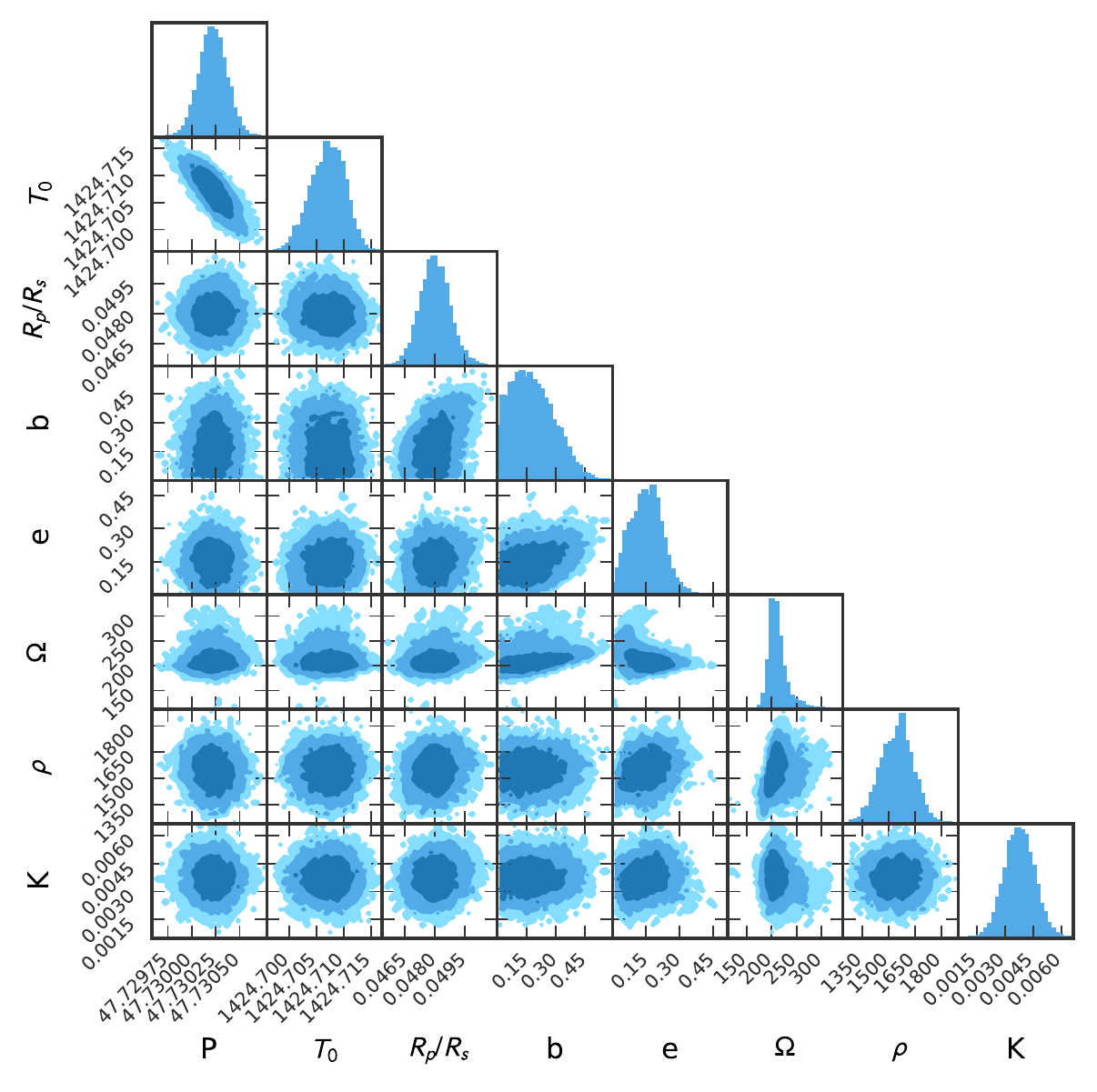}
    \caption{Posterior distributions of fitted parameters for \ticA\,b along with the stellar density ($\rm \rho$) and radial velocity semi-amplitude (K).}
    \label{fig:corner_1240}
  \end{figure*}

\end{appendix}

\vfill
\eject
\end{document}